# Multiple scattering theory for strong scattering heterogeneous elastic continua with triaxial inhomogeneities: theoretical fundamentals and applications

Huijing He[†]

*Department of Earth and Planetary Sciences, University of California, Santa Cruz,
Santa Cruz, CA 95064, USA*

[†] he.hui.jing@hotmail.com

## Abstract

The geometry of mesoscopic inhomogeneities plays an important role in determining the macroscopic propagation behaviors of elastic waves in a heterogeneous medium. Non-equiaxed inhomogeneities can lead to anisotropic wave velocity and attenuation. Developing an accurate scattering theory to describe the quantitative relation between the microstructure features and wave propagation parameters is of fundamental importance for seismology and ultrasonic nondestructive characterization. This work presents a multiple scattering theory for strongly scattering elastic media with general tri-axial heterogeneities. A closed analytical expression of the shape-dependent singularity of the anisotropic Green's tensor for the homogeneous reference medium is derived by introducing a proper non-orthogonal ellipsoidal coordinate. Renormalized Dyson's equation for the coherent wave field is then derived with the help of Feynman's diagram technique and the first-order-smoothing approximation. The exact dispersion curves and the inverse Q-factors of coherent waves in several representative medium models for the heterogeneous lithosphere are calculated numerically. Numerical results for small-scale heterogeneities with the aspect ratio varying from 1 to 7 show satisfactory agreement with those obtained from real earthquakes. The results for velocity dispersion give rise to a novel explanation to the formation mechanism of different seismic phases. The new model has potential applications in seismology and ultrasonic microstructure characterization.

**Key Words** scattering, heterogeneity, anisotropy, elastic waves, dispersion, attenuation, Mohorovičić discontinuity

## I. Introduction

Heterogeneous materials with non-equiaxed inhomogeneities widely exist in the nature. The velocity well-log data from the German Continental Deep Drilling Project (KTB) reveal that the aspect ration of the crustal heterogeneities in that area is about 1.8 [1]. P-wave and S-wave images obtained using travel-time tomography shows the aspect ratio of the crustal heterogeneities in southeast China varies from 3 to 7, or even larger [2]. Polycrystalline medium model of the Earth inner core which is assumed to be composed of perfectly aligned ellipsoidal grains is introduced to analyze the attenuation anisotropy in [3]. Non-equiaxed microstructures are also formed in titanium alloys and nickel-based super alloys as a result of plastic deformation caused by certain thermomechanical processing procedures such as rolling [4]. Solid-state phase transfer caused by heating and annealing often result in the formation of lamellar microstructures in titanium alloys [5-6]. Elastic waves propagating in random media with non-equiaxed heterogeneities exhibit complex behaviors, for example, the wave velocity and attenuation show complicated pattern depending on the grain shape and frequency. Conversely, the complicated wave field carries rich information about the microstructures, and provides the possibility to quantitative characterize or reconstruct the microstructures. Multiple scattering theory plays a key role in correlating the wavefield characteristic parameters with the microstructure features. It is also the theoretical basis for the development of grain noise measurement models and microstructure inversion model. Developing multiple scattering models for non-equiaxed microstructures has drawn broad attentions of researchers from the electromagnetic scattering, seismological and ultrasonics communities. Calvet and Margerin [3] generalized Weaver's weak scattering model for equiaxed grains to the case of general ellipsoidal grains. They obtained the approximate dispersion and attenuation properties of polycrystals with elongated grains using the spectral method. The results are used to explain the anisotropy of velocity and attenuation observed seismic waves from the Earth inner core. Yang et. al. [7] generalized Karal-Keller's model for equiaxed grain polycrystals [8-9] to those with elongated microstructures. The theoretical foundation of all these models are all based on the weak scattering approximation, which restricts their applications to materials with small density and elastic modulus perturbation. Development of strong scattering theory for non-equiaxed heterogeneous materials was first attempted by Kong and Tsang [10-11]. They systematically studied the calculation of the singularity tensors for dielectric materials with random ellipsoidal and cylindrical scatterers. Zhuck and Lakhtakia [12] studied the singularity of elastodynamic Green's function and derived the renormalization scheme for strong-property-fluctuation elastic media. Enlightened by the pioneering work by Kong and Tsang, and Zhuck, He [13-14] developed a strong-property-fluctuation multiple scattering theory and obtained the exact solution to the dispersion equations for both two-phase locally isotropic elastic materials and polycrystalline materials. He and Kube [15] systematically compared the performance of the weak scattering models and the strong scattering models, pointing out that for materials with simultaneous density and elastic modulus perturbations, the weak scattering theory gives unstable results for velocity and attenuation, while the strong scattering theory is capable to provide robust predictions. However, all these works only considered heterogeneous materials with equiaxed heterogeneities. Although non-equiaxed heterogeneous materials have drawn broad and long-lasting interests in ultrasonic NDE and seismology, a multiple scattering theory for strong scattering materials is still missing. In this work, a novel multiple scattering theory for the strong scattering medium with general tri-axial ellipsoidal heterogeneities is developed. In Sec. II, the general theory for anisotropic Green's functions, the singularity of Green's tensor, and the renormalized Dyson's equation are discussed in detail. Section III gives concrete numerical examples for a series of representative material models of the heterogeneous lithosphere are given. The velocity and attenuation of coherent seismic waves are solved from the dispersion equation. The implications of the new results and possible future developments of the new theory are discussed in Sec. IV. In the Appendix we give the closed analytical expression of the singularity part of Green's tensor for a general tri-axial heterogeneous medium.

## II. Theoretical fundamentals

In this section we present the detailed development of the multiple scattering theory for strongly scattering elastic media with tri-axial heterogeneities. Starting from the elastodynamic equations and Green's function for general anisotropic materials, an integral representation for the perturbed wavefield is formulated. After introducing a proper non-orthogonal ellipsoidal coordinate system for Green's function in the frequency-wavenumber domain, we obtained the explicit integral expression of the shape-dependent singularity tensor. Dyson's equation for the renormalized field is then derived with the help of Feynman's diagram technique under the first-order-smoothing approximation. Finally, the dispersion equation for the coherent waves are obtained by applying Fourier transform to Dyson's equation.

### II-1. Green's function and the integral representation

Heterogeneous media with non-equiaxed heterogeneities exhibit obvious anisotropic elastic properties. Consequently, the proper choice of the homogeneous reference medium is an anisotropic material with certain degree of symmetry, which is dictated by the statistical characteristics of the random inhomogeneities. The integral representation of the scattering wavefield relies on the Green's function of the reference medium, so we first carry out a detailed analysis of the Green's function of a general anisotropic homogeneous medium. Suppose a time-harmonic unit concentrated force $\mathbf{F}$ is applied to an infinite homogeneous medium at point $\mathbf{x}'$ along the direction $\mathbf{e}_\alpha$, the displacement field is defined as the Green's function, governed by the elastodynamic equation

$$c_{ijkl}G_{k\alpha,lj} + \rho\omega^2 G_{i\alpha} + Fa_{i\alpha}\delta(\mathbf{x}-\mathbf{x}') = 0, \tag{1}$$

where $c_{ijkl}$ and $\rho$ are the elastic stiffness tensor and the density, respectively. $F$ is the magnitude of the force, we assume $F$=1 Newton. The Cartesian tensor notation and the Einstein summation convention are used throughout this work, and an index following a comma implies partial derivative with respect to the corresponding spatial coordinate. $O\text{-}x_i x_j x_k$ is the coordinate system defined for the wavefield, and $O' - x'_\alpha x'_\beta x'_\gamma$ is the coordinate system for the source distribution. $a_{i\alpha}$ is the direction cosine defined by $a_{i\alpha} = \cos(\mathbf{e}_\alpha, \mathbf{e}_i)$. The time-harmonic factor $e^{-i\omega t}$ is omitted in Eq. (1) and it will be added when necessary.

To facilitate the analysis of the Green's function in the frequency-wavenumber domain, we introduce the spatial Fourier transform pair defined by

$$\begin{aligned}\tilde{G}_{i\alpha}(\mathbf{k},\mathbf{x}',\omega) &= \iiint_{V(\mathbf{x})} G_{i\alpha}(\mathbf{x},\mathbf{x}',\omega)e^{-i\mathbf{k}\cdot\mathbf{x}}d^3\mathbf{x}, \\ G_{i\alpha}(\mathbf{x},\mathbf{x}',\omega) &= \frac{1}{(2\pi)^3}\iiint_{V(\mathbf{k})}\tilde{G}_{i\alpha}(\mathbf{k},\mathbf{x}',\omega)e^{i\mathbf{k}\cdot\mathbf{x}}d^3\mathbf{k}.\end{aligned} \tag{2}$$

Under the assumption that the medium has a small damping and considering the Sommerfeld radiation condition, the integral of the Green's function and its derivatives over a closed surface enclosing the source point but infinitely far from the source region vanishes, i.e.

$$\oiint_S G_{i\alpha}(\mathbf{x},\mathbf{x}',\omega)e^{-i\mathbf{k}\cdot\mathbf{x}}N_j dS = 0, \text{ as } |\mathbf{x}-\mathbf{x}'|\to\infty, \text{ Im}(k)<0, \tag{3}$$

thus, the Fourier transform of the first and second order derivatives of Green's function are

$$\mathscr{F}\left(\frac{\partial G_{i\alpha}}{\partial x_j}\right) = ik_j \tilde{G}_{i\alpha}(\mathbf{k},\mathbf{x}',\omega), \quad \mathscr{F}\left(\frac{\partial^2 G_{i\alpha}}{\partial x_j \partial x_l}\right) = -k_j k_l \tilde{G}_{i\alpha}(\mathbf{k},\mathbf{x}',\omega). \tag{4}$$

Applying Fourier transform to Eq. (1) and considering the relations in Eq. (4), we obtain

$$(c_{ijkl}k_l k_j - \rho\omega^2 \delta_{ik})\tilde{G}_{k\alpha}(\mathbf{k},\mathbf{x}',\omega) = F a_{i\alpha} e^{-i\mathbf{k}\cdot\mathbf{x}'}, \tag{5}$$

In subsequent discussion, we assume that the coordinate system for the source coincides with that for the wavefield, so we have $a_{i\alpha} = \delta_{i\alpha}$.

In Eq. (5) we can introduce the Christoffel tensor defined by $\Gamma_{ik} = c_{ijkl}k_l k_j$, which is a symmetric, positive definite tensor. For a general anisotropic media, the Christoffel tensor takes the form

$$\Gamma_{ij} = \begin{bmatrix} \Gamma_{11} & \Gamma_{12} & \Gamma_{13} \\ \Gamma_{12} & \Gamma_{22} & \Gamma_{23} \\ \Gamma_{13} & \Gamma_{23} & \Gamma_{33} \end{bmatrix}, \tag{6}$$

where its components are given by

$$\Gamma_{11} = c_{11}k_1^2 + c_{66}k_2^2 + c_{55}k_3^2 + 2c_{16}k_1 k_2 + 2c_{15}k_1 k_3 + 2c_{56}k_2 k_3, \tag{7a}$$

$$\Gamma_{12} = c_{16}k_1^2 + c_{26}k_2^2 + c_{45}k_3^2 + (c_{12}+c_{66})k_1 k_2 + (c_{14}+c_{56})k_1 k_3 + (c_{46}+c_{25})k_2 k_3, \tag{7b}$$

$$\Gamma_{13} = c_{15}k_1^2 + c_{46}k_2^2 + c_{35}k_3^2 + (c_{14}+c_{56})k_1 k_2 + (c_{13}+c_{55})k_1 k_3 + (c_{36}+c_{45})k_2 k_3, \tag{7c}$$

$$\Gamma_{22} = c_{66}k_1^2 + c_{22}k_2^2 + c_{44}k_3^2 + 2c_{26}k_1 k_2 + 2c_{46}k_1 k_3 + 2c_{24}k_2 k_3, \tag{7d}$$

$$\Gamma_{23} = c_{56}k_1^2 + c_{24}k_2^2 + c_{34}k_3^2 + (c_{46}+c_{25})k_1 k_2 + (c_{36}+c_{45})k_1 k_3 + (c_{23}+c_{44})k_2 k_3, \tag{7e}$$

$$\Gamma_{33} = c_{55}k_1^2 + c_{44}k_2^2 + c_{33}k_3^2 + 2c_{45}k_1 k_2 + 2c_{35}k_1 k_3 + 2c_{34}k_2 k_3. \tag{7f}$$

Green's tensor in the frequency-wavenumber domain can be expressed in terms of the Christoffel tensor as

$$\begin{bmatrix} \tilde{G}_{11} & \tilde{G}_{12} & \tilde{G}_{13} \\ \tilde{G}_{12} & \tilde{G}_{22} & \tilde{G}_{23} \\ \tilde{G}_{13} & \tilde{G}_{23} & \tilde{G}_{33} \end{bmatrix} = e^{-i\mathbf{k}\cdot\mathbf{x}'} \begin{bmatrix} \Gamma_{11}-\rho\omega^2 & \Gamma_{12} & \Gamma_{13} \\ \Gamma_{12} & \Gamma_{22}-\rho\omega^2 & \Gamma_{23} \\ \Gamma_{13} & \Gamma_{23} & \Gamma_{33}-\rho\omega^2 \end{bmatrix}^{-1}, \tag{8}$$

To obtain an explicit expression of Green's tensor, we further need the eigenspace expansion of the Christoffel tensor. Introduce a modified Christoffel tensor $\bar{\Gamma}_{ij}$ defined by

$$\bar{\Gamma}_{ij} = \frac{\Gamma_{ij}}{k^2}, \tag{9}$$

where $k$ is the magnitude of the wavevector $\mathbf{k}$.

As a result of the positive definiteness of the Christoffel tensor, its three eigenvalues are all positive, which are denoted by

$$\lambda_1 = \hat{c}_1, \ \lambda_2 = \hat{c}_2, \lambda_3 = \hat{c}_3, \ \hat{c}_1 > \hat{c}_2 \geq \hat{c}_3 > 0, \tag{10}$$

the corresponding eigenvectors are

$$\mathbf{q}^{(1)} = [q_1^{(1)} \ q_2^{(1)} \ q_3^{(1)}]^T, \ \mathbf{q}^{(2)} = [q_1^{(2)} \ q_2^{(2)} \ q_3^{(2)}]^T, \ \mathbf{q}^{(3)} = [q_1^{(3)} \ q_2^{(3)} \ q_3^{(3)}]^T. \tag{11}$$

The three eigenvectors constitute an orthonormal tetrad, and the orthogonal transfer matrix $\mathbf{Q}$ is given by

$$\mathbf{Q} = [\mathbf{q}^{(1)} \ \mathbf{q}^{(2)} \ \mathbf{q}^{(3)}]. \tag{12}$$

The eigenspace expansion of the modified Christoffel tensor can be written as

$$\bar{\Gamma} = \mathbf{Q}\Lambda\mathbf{Q}^T, \tag{13}$$

where $\mathbf{\Lambda} = diag(\hat{c}_1, \hat{c}_2, \hat{c}_3)$.

Finally, we obtain the eigenspace expansion of Green's function

$$\tilde{\mathbf{G}}(\mathbf{k}, \mathbf{x}', \omega) = \sum_{n=1}^{3} \frac{F e^{-i\mathbf{k} \cdot \mathbf{x}'}}{\hat{c}_n k^2 - \rho \omega^2} \mathbf{q}^{(n)} \otimes \mathbf{q}^{(n)}, \tag{14}$$

where $\otimes$ denotes tensor product of the related vectors.

Green's function in the spatial-frequency domain is obtained by inverse Fourier transform

$$\mathbf{G}(\mathbf{x}, \mathbf{x}', \omega) = \frac{1}{(2\pi)^3} \iiint_{V(\mathbf{k})} \sum_{n=1}^{3} \frac{F e^{i\mathbf{k} \cdot (\mathbf{x}-\mathbf{x}')}}{\hat{c}_n k^2 - \rho \omega^2} \mathbf{q}^{(n)} \otimes \mathbf{q}^{(n)} d^3\mathbf{k}, \tag{15}$$

Later we will consider a specific class of heterogeneous media in which all the tri-axial inhomogeneities have the same aspect ratio and the major axes are perfectly aligned. The homogeneous reference medium for this type of materials has orthogonal symmetry, so the elastic stiffness tensor of the reference medium is given by

$$c_{ij} = \begin{bmatrix} c_{11} & c_{12} & c_{13} & 0 & 0 & 0 \\ c_{12} & c_{22} & c_{23} & 0 & 0 & 0 \\ c_{13} & c_{23} & c_{33} & 0 & 0 & 0 \\ 0 & 0 & 0 & c_{44} & 0 & 0 \\ 0 & 0 & 0 & 0 & c_{55} & 0 \\ 0 & 0 & 0 & 0 & 0 & c_{66} \end{bmatrix}. \tag{16}$$

The corresponding Christoffel tensor is given by

$$\Gamma_{11} = c_{11}k_1^2 + c_{66}k_2^2 + c_{55}k_3^2, \quad \Gamma_{22} = c_{66}k_1^2 + c_{22}k_2^2 + c_{44}k_3^2, \quad \Gamma_{33} = c_{55}k_1^2 + c_{44}k_2^2 + c_{33}k_3^2, \tag{17a}$$

$$\Gamma_{12} = (c_{12} + c_{66})k_1 k_2, \quad \Gamma_{13} = (c_{13} + c_{55})k_1 k_3, \quad \Gamma_{23} = (c_{23} + c_{44})k_2 k_3. \tag{17b}$$

For static problems, the static Green's function takes the form

$$\tilde{G}_{11}^s(\mathbf{k}) = \frac{1}{\Delta_s}(\Gamma_{22}\Gamma_{33} - \Gamma_{23}^2), \quad \tilde{G}_{22}^s(\mathbf{k}) = \frac{1}{\Delta_s}(\Gamma_{11}\Gamma_{33} - \Gamma_{13}^2), \quad \tilde{G}_{33}^s(\mathbf{k}) = \frac{1}{\Delta_s}(\Gamma_{11}\Gamma_{22} - \Gamma_{12}^2),$$

$$\tilde{G}_{12}^s(\mathbf{k}) = \frac{1}{\Delta_s}(\Gamma_{13}\Gamma_{23} - \Gamma_{12}\Gamma_{33}), \quad \tilde{G}_{23}^s(\mathbf{k}) = \frac{1}{\Delta_s}(\Gamma_{12}\Gamma_{13} - \Gamma_{23}\Gamma_{11}), \quad \tilde{G}_{13}^s(\mathbf{k}) = \frac{1}{\Delta_s}(\Gamma_{12}\Gamma_{23} - \Gamma_{13}\Gamma_{22}), \tag{18}$$

where

$$\Delta_s = \det \begin{bmatrix} \Gamma_{11} & \Gamma_{12} & \Gamma_{13} \\ \Gamma_{12} & \Gamma_{22} & \Gamma_{23} \\ \Gamma_{13} & \Gamma_{23} & \Gamma_{33} \end{bmatrix}. \tag{19}$$

The integral representation of the scattering wavefield for a general inhomogeneous medium can be expressed in terms of Green's function of the homogeneous reference medium as [13]

$$\begin{bmatrix} G_{\beta''a'}(\mathbf{x}'', \mathbf{x}') \\ \varepsilon_{\alpha''\beta''a'}(\mathbf{x}'', \mathbf{x}') \end{bmatrix} = \begin{bmatrix} G_{\beta''a'}^0(\mathbf{x}'', \mathbf{x}') \\ \varepsilon_{\alpha''\beta''a'}^0(\mathbf{x}'', \mathbf{x}') \end{bmatrix} + \int_V \begin{bmatrix} G_{\beta''i''}^0(\mathbf{x}'', \mathbf{x}) & \varepsilon_{\beta''i''j''}^0(\mathbf{x}'', \mathbf{x}) \\ \varepsilon_{\alpha''\beta''i''}^0(\mathbf{x}'', \mathbf{x}) & E_{\alpha''\beta''i''j''}^0(\mathbf{x}'', \mathbf{x}) \end{bmatrix} \begin{bmatrix} \delta\rho(\mathbf{x})\omega^2 \delta_{ij} & 0 \\ 0 & \delta c_{ijkl}(\mathbf{x}) \end{bmatrix} \begin{bmatrix} G_{ja'}(\mathbf{x}, \mathbf{x}') \\ \varepsilon_{kla'}(\mathbf{x}, \mathbf{x}') \end{bmatrix} dV, \tag{20}$$

where $G_{\beta''a'}(\mathbf{x}'', \mathbf{x}')$ and $\varepsilon_{\alpha''\beta''a'}(\mathbf{x}'', \mathbf{x}')$ are the displacement and strain Green's function of the heterogeneous medium, $G_{\beta''i''}^0(\mathbf{x}'', \mathbf{x})$, $\varepsilon_{\beta''i''j''}^0(\mathbf{x}'', \mathbf{x})$ and $E_{\alpha''\beta''i''j''}^0(\mathbf{x}'', \mathbf{x})$ are the components of the Green's tensor of the homogeneous reference medium. $\delta c_{ijkl}(\mathbf{x})$ and $\delta\rho(\mathbf{x})$ are the perturbation of the elastic stiffness and density, respectively. The detailed derivation and the explicit expression of the homogeneous Green's tensor are given in [13].

Eq. (20) can be expressed more compactly in a matrix form,

$$\Psi(\mathbf{x}''-\mathbf{x}') = \Psi^0(\mathbf{x}''-\mathbf{x}') + \iiint_{V(\mathbf{x})} \Gamma(\mathbf{x}''-\mathbf{x})\Pi(\mathbf{x})\Psi(\mathbf{x}-\mathbf{x}')d^3\mathbf{x} \ . \tag{21}$$

The detailed expression of the symbols can be found in [13].

## II-2. Singularity of Green's tensor

It has been pointed in [13] that the Green's tensor components involving second order derivative of the displacement Green's function have $\delta$ singularities, and the proper calculation of the integral in Eq. (21) require the introduction of the shape-dependent principle value of the Green's tensor. It is well known that for a heterogeneous medium with equiaxed inhomogeneities, the reference medium is isotropic, for which the explicit expression of Green's tensor in the spatial-frequency domain is available. Consequently, the singularity part can be calculated directly from the spatial domain Green's function, as discussed in Method 1 in the appendix of [13]. For heterogeneous media with non-equiaxed heterogeneities, the reference medium is anisotropic. An inspection of Eq. (15) tells us that Green's function in the spatial-frequency domain cannot be calculated explicitly. Consequently, calculating the singularity tensor by using the spatial domain Green's functions is impossible. Attempts to calculate the singularity part by using the frequency-wavenumber domain Green's function also fail because the dependence of eigenvalues on the wavevector direction obstructs the calculation of the integrals in Eq. (A42) in [13]. As a result, Method 2 introduced in the appendix of [13] is failed either. Therefore, Method 3 in the Appendix of [13] is the only approach to calculate the singularity of the anisotropic Green's function.

According to the analysis in [13], the homogeneous reference medium should be chosen so that the convolution integral of the shape-dependent principle value of the Green's tensor with the spectral function of the spatial correlation functions vanishes when the frequency approaches zero, i.e.,

$$\lim_{\omega \to 0} \mathscr{F}[P.S.E^0_{\alpha ij\beta}(\mathbf{x}-\mathbf{x}',\omega)P(\mathbf{x}-\mathbf{x}')] = 0 \ . \tag{22}$$

where $P(\mathbf{x}-\mathbf{x}')$ is the spatial correlation function of a random medium.

Considering the definition of the shape-dependent principle value

$$P.S.E^0_{\alpha ij\beta}(\mathbf{x}-\mathbf{x}',\omega) = E^0_{\alpha ij\beta}(\mathbf{x}-\mathbf{x}',\omega) + S_{\alpha ij\beta}\delta(\mathbf{x}-\mathbf{x}') \ , \tag{23}$$

we have

$$\begin{aligned}
&\mathscr{F}[P.S.E^0_{\alpha ij\beta}(\mathbf{x}-\mathbf{x}',\omega)P(\mathbf{x}-\mathbf{x}')] \\
&= \iiint_{V(\mathbf{x})} [E^0_{\alpha ij\beta}(\mathbf{x}-\mathbf{x}',\omega) + S_{\alpha ij\beta}\delta(\mathbf{x}-\mathbf{x}')]P(\mathbf{x}-\mathbf{x}')e^{-i\mathbf{k}\cdot\mathbf{x}}d^3\mathbf{x} \\
&= S_{\alpha ij\beta}P(0)e^{-i\mathbf{k}\cdot\mathbf{x}'} + \iiint_{V(\mathbf{x})} E^0_{\alpha ij\beta}(\mathbf{x}-\mathbf{x}',\omega)P(\mathbf{x}-\mathbf{x}')e^{-i\mathbf{k}\cdot\mathbf{x}}d^3\mathbf{x} \\
&= S_{\alpha ij\beta}P(0)e^{-i\mathbf{k}\cdot\mathbf{x}'} + \frac{1}{8\pi^3}\iiint_{V(\mathbf{s})} \tilde{E}^0_{\alpha ij\beta}(\mathbf{s},\omega)\tilde{P}(\mathbf{k}-\mathbf{s})d^3\mathbf{s},
\end{aligned} \tag{24}$$

The zero-frequency limit of Eq. (24) is obtained by setting $\omega \to 0$. Inserting Eq. (24) into Eq. (22) and taking the zero-frequency limit, we obtain

$$S_{\alpha ij\beta} + \frac{1}{8\pi^3}\lim_{\omega \to 0}\iiint_{V(\mathbf{s})} \tilde{E}^0_{\alpha ij\beta}(\mathbf{s},\omega)\tilde{P}(-\mathbf{s})d^3\mathbf{s} = 0, \tag{25}$$

where $P(0) = 1$ is used.

Finally, we obtain the analytic expression of the singularity

$$S_{\alpha ij\beta} = -\frac{1}{8\pi^3}\lim_{\omega \to 0}\iiint_{V(\mathbf{s})} \tilde{E}^0_{\alpha ij\beta}(\mathbf{s},\omega)\tilde{P}(-\mathbf{s})d^3\mathbf{s}, \tag{26}$$

where

$$\tilde{E}^0_{\alpha ij\beta}(\mathbf{k},\omega) = -\frac{1}{4}[k_j k_\beta \tilde{G}^0_{\alpha i}(\mathbf{k},\omega) + k_i k_\beta \tilde{G}^0_{\alpha j}(\mathbf{k},\omega) + k_j k_\alpha G^0_{\beta i}(\mathbf{k},\omega) + k_i k_\alpha G^0_{\beta j}(\mathbf{k},\omega)]. \tag{27}$$

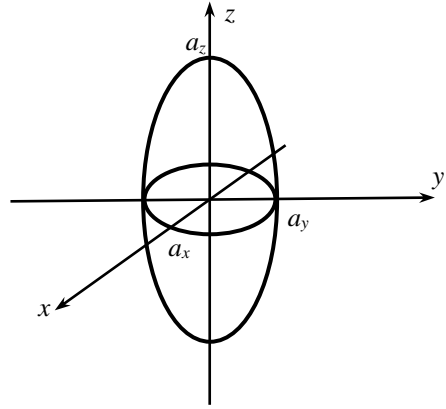

Figure 1. Representative ellipsoid of a heterogeneous medium with tri-axial inhomogeneities

Geometric anisotropy of non-equiaxed inhomogeneities are conveniently described by the anisotropic two-point correlation function. For tri-axial inhomogeneities, the spatial correlation function is given in [3]

$$P(\mathbf{r}) = \exp\left[-\sqrt{\frac{r_1^2}{a_x^2} + \frac{r_2^2}{a_y^2} + \frac{r_3^2}{a_z^2}}\right], \qquad \mathbf{r} = \mathbf{x} - \mathbf{x}', \tag{28}$$

where $a_x$, $a_y$ and $a_z$ are the semi-axes of the representative ellipsoid as shown in Fig. 1. The representative ellipsoid is also the level set of the two-point correlation function at which the possibility distribution function of the grain size assumes its maximum value.

The power spectrum density of the random medium, i.e., the Fourier transform of the correlation function takes the form

$$\tilde{P}(\mathbf{k}) = \frac{8\pi a_x a_y a_z}{(1 + k_x^2 a_x^2 + k_y^2 a_y^2 + k_z^2 a_z^2)^2}. \tag{29}$$

Therefore, the singularity is given by [13]

$$S_{\alpha ij\beta} = \frac{a_x a_y a_z}{4\pi^2} \iiint_{V(\mathbf{s})} [s_j s_\beta \tilde{G}^0_{\alpha i}(\mathbf{s},\omega) + s_i s_\beta \tilde{G}^0_{\alpha j}(\mathbf{s},\omega) + s_j s_\alpha G^0_{\beta i}(\mathbf{s},\omega) + s_i s_\alpha G^0_{\beta j}(\mathbf{s},\omega)] \frac{d^3\mathbf{s}}{(1 + s_x^2 a^2 + s_y^2 b^2 + s_z^2 c^2)^2}. \tag{30}$$

The detailed calculation of the singularity tensor is presented in Appendix A.

With the help of the shape-dependent principle value, we obtain the correct definition of the integrals appeared in the representation of the scattering wavefield

$$\mathbf{\Psi}(\mathbf{x}'' - \mathbf{x}') = \mathbf{\Psi}^0(\mathbf{x}'' - \mathbf{x}') + \iiint_{V(\mathbf{x})}[P.S.\mathbf{\Gamma}(\mathbf{x}'' - \mathbf{x}) - \mathbf{S}\delta(\mathbf{x}'' - \mathbf{x})]\mathbf{\Pi}(\mathbf{x})\mathbf{\Psi}(\mathbf{x} - \mathbf{x}')d^3\mathbf{x}. \tag{31}$$

Using the definition of Dirac-$\delta$ function, we get

$$\mathbf{\Psi}(\mathbf{x}'' - \mathbf{x}') = \mathbf{\Psi}^0(\mathbf{x}'' - \mathbf{x}') + \iiint_{V(\mathbf{x})} P.S.\mathbf{\Gamma}(\mathbf{x}'' - \mathbf{x})d^3\mathbf{x} - \mathbf{S}\mathbf{\Pi}(\mathbf{x}'')\mathbf{\Psi}(\mathbf{x}'' - \mathbf{x}'). \tag{32}$$

Introducing the renormalized variable

$$\mathbf{\Phi}(\mathbf{x}'' - \mathbf{x}') = \mathbf{\Psi}(\mathbf{x}'' - \mathbf{x}') + \mathbf{S}\mathbf{\Pi}(\mathbf{x}'')\mathbf{\Psi}(\mathbf{x}'' - \mathbf{x}'), \tag{33}$$

we can rewrite the integral equation as

$$\mathbf{\Phi}(\mathbf{x}'' - \mathbf{x}') = \mathbf{\Psi}^0(\mathbf{x}'' - \mathbf{x}') + \iiint_{V(\mathbf{x})} P.S.\mathbf{\Gamma}(\mathbf{x}'' - \mathbf{x})\mathbf{\Xi}(\mathbf{x})\mathbf{\Phi}(\mathbf{x} - \mathbf{x}')d^3\mathbf{x}, \tag{34}$$

where $\mathbf{\Xi}(\mathbf{x})$ is the renormalized property perturbation matrix

$$\Xi(\mathbf{x}) = \Pi(\mathbf{x})[\mathbf{I} + \mathbf{S}\Pi(\mathbf{x})]^{-1}. \tag{35}$$

For the heterogeneous medium with triaxial inhomogeneities, $\Xi(\mathbf{x})$ takes the form

$$\Xi_{ij} = \begin{bmatrix} \Xi_{11} & \Xi_{12} & \Xi_{13} & 0 & 0 & 0 \\ \Xi_{12} & \Xi_{22} & \Xi_{23} & 0 & 0 & 0 \\ \Xi_{13} & \Xi_{23} & \Xi_{33} & 0 & 0 & 0 \\ 0 & 0 & 0 & \Xi_{44} & 0 & 0 \\ 0 & 0 & 0 & 0 & \Xi_{55} & 0 \\ 0 & 0 & 0 & 0 & 0 & \Xi_{66} \end{bmatrix}, \tag{36}$$

where

$$\Xi_{11} = \frac{1}{\Delta_\Xi}[\delta c_{11} + S_{22}(\delta c_{11}\delta c_{22} - \delta c_{12}^2) + S_{33}(\delta c_{11}\delta c_{33} - \delta c_{13}^2) + 2S_{23}(\delta c_{11}\delta c_{23} - \delta c_{12}\delta c_{13}) + \\ (S_{23}^2 - S_{22}S_{33})(\delta c_{11}\delta c_{23}^2 + \delta c_{22}\delta c_{13}^2 + \delta c_{33}\delta c_{12}^2 - 2\delta c_{12}\delta c_{13}\delta c_{23} - \delta c_{11}\delta c_{22}\delta c_{33})], \tag{37a}$$

$$\Xi_{22} = \frac{1}{\Delta_\Xi}[\delta c_{22} + S_{11}(\delta c_{11}\delta c_{22} - \delta c_{12}^2) + S_{33}(\delta c_{22}\delta c_{33} - \delta c_{23}^2) + 2S_{13}(\delta c_{22}\delta c_{13} - \delta c_{12}\delta c_{23}) + \\ (S_{13}^2 - S_{11}S_{33})(\delta c_{11}\delta c_{23}^2 + \delta c_{22}\delta c_{13}^2 + \delta c_{33}\delta c_{12}^2 - 2\delta c_{12}\delta c_{13}\delta c_{23} - \delta c_{11}\delta c_{22}\delta c_{33})], \tag{37b}$$

$$\Xi_{33} = \frac{1}{\Delta_\Xi}[\delta c_{33} + S_{11}(\delta c_{11}\delta c_{33} - \delta c_{13}^2) + S_{22}(\delta c_{22}\delta c_{33} - \delta c_{23}^2) + 2S_{12}(\delta c_{33}\delta c_{12} - \delta c_{13}\delta c_{23}) + \\ (S_{12}^2 - S_{11}S_{22})(\delta c_{11}\delta c_{23}^2 + \delta c_{22}\delta c_{13}^2 + \delta c_{33}\delta c_{12}^2 - 2\delta c_{12}\delta c_{13}\delta c_{23} - \delta c_{11}\delta c_{22}\delta c_{33})], \tag{37c}$$

$$\Xi_{12} = \frac{1}{\Delta_\Xi}[\delta c_{12} + S_{12}(\delta c_{12}^2 - \delta c_{11}\delta c_{22}) + S_{13}(\delta c_{12}\delta c_{13} - \delta c_{11}\delta c_{23}) + S_{23}(\delta c_{23}\delta c_{12} - \delta c_{13}\delta c_{22}) + S_{33}(\delta c_{33}\delta c_{12} - \delta c_{13}\delta c_{23}) + \\ (S_{12}S_{33} - S_{13}S_{23})(\delta c_{11}\delta c_{23}^2 + \delta c_{22}\delta c_{13}^2 + \delta c_{33}\delta c_{12}^2 - 2\delta c_{12}\delta c_{13}\delta c_{23} - \delta c_{11}\delta c_{22}\delta c_{33})], \tag{37d}$$

$$\Xi_{13} = \frac{1}{\Delta_\Xi}[\delta c_{13} + S_{13}(\delta c_{13}^2 - \delta c_{11}\delta c_{33}) + S_{12}(\delta c_{12}\delta c_{13} - \delta c_{11}\delta c_{23}) + S_{22}(\delta c_{13}\delta c_{22} - \delta c_{12}\delta c_{23}) + S_{23}(\delta c_{13}\delta c_{23} - \delta c_{12}\delta c_{33}) + \\ (S_{13}S_{22} - S_{12}S_{23})(\delta c_{11}\delta c_{23}^2 + \delta c_{22}\delta c_{13}^2 + \delta c_{33}\delta c_{12}^2 - 2\delta c_{12}\delta c_{13}\delta c_{23} - \delta c_{11}\delta c_{22}\delta c_{33})], \tag{37e}$$

$$\Xi_{23} = \frac{1}{\Delta_\Xi}[\delta c_{23} + S_{11}(\delta c_{11}\delta c_{23} - \delta c_{12}\delta c_{13}) + S_{12}(\delta c_{12}\delta c_{23} - \delta c_{13}\delta c_{22}) + S_{13}(\delta c_{13}\delta c_{23} - \delta c_{12}\delta c_{33}) + S_{23}(\delta c_{23}^2 - \delta c_{22}\delta c_{33}) + \\ (S_{23}S_{11} - S_{12}S_{13})(\delta c_{11}\delta c_{23}^2 + \delta c_{22}\delta c_{13}^2 + \delta c_{33}\delta c_{12}^2 - 2\delta c_{12}\delta c_{13}\delta c_{23} - \delta c_{11}\delta c_{22}\delta c_{33})], \tag{37f}$$

$$\Xi_{44} = \frac{\delta c_{44}}{1 + 4S_{44}\delta c_{44}}, \quad \Xi_{55} = \frac{\delta c_{55}}{1 + 4S_{55}\delta c_{55}}, \quad \Xi_{66} = \frac{\delta c_{66}}{1 + 4S_{66}\delta c_{66}}, \tag{37g}$$

$$\Delta_\Xi = 1 + S_{11}\delta c_{11} + S_{22}\delta c_{22} + S_{33}\delta c_{33} + 2S_{12}\delta c_{12} + 2S_{13}\delta c_{13} + 2S_{23}\delta c_{23} + (\delta c_{11}\delta c_{22} - \delta c_{12}^2)(S_{11}S_{22} - S_{12}^2) + \\ (\delta c_{11}\delta c_{33} - \delta c_{13}^2)(S_{11}S_{33} - S_{13}^2) + (\delta c_{22}\delta c_{33} - \delta c_{23}^2)(S_{22}S_{33} - S_{23}^2) + 2(\delta c_{11}\delta c_{23} - \delta c_{12}\delta c_{13})(S_{11}S_{23} - S_{12}S_{13}) \\ + 2(\delta c_{22}\delta c_{13} - \delta c_{12}\delta c_{23})(S_{22}S_{13} - S_{12}S_{23}) + 2(\delta c_{33}\delta c_{12} - \delta c_{13}\delta c_{23})(S_{33}S_{12} - S_{13}S_{23}) \\ + (\delta c_{11}\delta c_{23}^2 + \delta c_{22}\delta c_{13}^2 + \delta c_{33}\delta c_{12}^2 - 2\delta c_{12}\delta c_{13}\delta c_{23} - \delta c_{11}\delta c_{22}\delta c_{33})(S_{11}S_{23}^2 + S_{22}S_{13}^2 + S_{33}S_{12}^2 - 2S_{12}S_{13}S_{23} - S_{11}S_{22}S_{33}). \tag{37h}$$

In this work, we consider a special class of random medium which is composed of two component phases, called Phase 1 and 2, both of which are isotropic elastic materials. The elastic stiffness perturbation can take two values only

$$\delta c_{ij}^{(1)} = \begin{bmatrix} \lambda_1 + 2\mu_1 - c_{11} & \lambda_1 - c_{12} & \lambda_1 - c_{13} & 0 & 0 & 0 \\ \lambda_1 - c_{12} & \lambda_1 + 2\mu_1 - c_{22} & \lambda_1 - c_{23} & 0 & 0 & 0 \\ \lambda_1 - c_{13} & \lambda_1 - c_{23} & \lambda_1 + 2\mu_1 - c_{33} & 0 & 0 & 0 \\ 0 & 0 & 0 & \mu_1 - c_{44} & 0 & 0 \\ 0 & 0 & 0 & 0 & \mu_1 - c_{55} & 0 \\ 0 & 0 & 0 & 0 & 0 & \mu_1 - c_{66} \end{bmatrix}, \text{ for Phase 1} \tag{38}$$

$$\delta c_{ij}^{(2)} = \begin{bmatrix} \lambda_2 + 2\mu_2 - c_{11} & \lambda_2 - c_{12} & \lambda_2 - c_{13} & 0 & 0 & 0 \\ \lambda_2 - c_{12} & \lambda_2 + 2\mu_2 - c_{22} & \lambda_2 - c_{23} & 0 & 0 & 0 \\ \lambda_2 - c_{13} & \lambda_2 - c_{23} & \lambda_2 + 2\mu_2 - c_{33} & 0 & 0 & 0 \\ 0 & 0 & 0 & \mu_2 - c_{44} & 0 & 0 \\ 0 & 0 & 0 & 0 & \mu_2 - c_{55} & 0 \\ 0 & 0 & 0 & 0 & 0 & \mu_2 - c_{66} \end{bmatrix}, \text{ for Phase 2} \quad (39)$$

Substitution of $\delta c_{ij}^{(1)}$ and $\delta c_{ij}^{(2)}$ into the expressions for $\Xi_{ij}$, we get $\Xi_{ij}^{(1)}$ and $\Xi_{ij}^{(2)}$, and the condition $\langle \Xi(\mathbf{x}) \rangle = 0$ implies

$$\rho = f_1 \rho_1 + f_2 \rho_2, \quad (40)$$

$$\begin{bmatrix} \Xi_{11}^{(1)} & \Xi_{12}^{(1)} & \Xi_{13}^{(1)} & 0 & 0 & 0 \\ \Xi_{12}^{(1)} & \Xi_{22}^{(1)} & \Xi_{23}^{(1)} & 0 & 0 & 0 \\ \Xi_{13}^{(1)} & \Xi_{23}^{(1)} & \Xi_{33}^{(1)} & 0 & 0 & 0 \\ 0 & 0 & 0 & \Xi_{44}^{(1)} & 0 & 0 \\ 0 & 0 & 0 & 0 & \Xi_{55}^{(1)} & 0 \\ 0 & 0 & 0 & 0 & 0 & \Xi_{66}^{(1)} \end{bmatrix} f_1 + \begin{bmatrix} \Xi_{11}^{(2)} & \Xi_{12}^{(2)} & \Xi_{13}^{(2)} & 0 & 0 & 0 \\ \Xi_{12}^{(2)} & \Xi_{22}^{(2)} & \Xi_{23}^{(2)} & 0 & 0 & 0 \\ \Xi_{13}^{(2)} & \Xi_{23}^{(2)} & \Xi_{33}^{(2)} & 0 & 0 & 0 \\ 0 & 0 & 0 & \Xi_{44}^{(2)} & 0 & 0 \\ 0 & 0 & 0 & 0 & \Xi_{55}^{(2)} & 0 \\ 0 & 0 & 0 & 0 & 0 & \Xi_{66}^{(2)} \end{bmatrix} f_2 = \mathbf{0}, \quad (41)$$

The material properties of the reference medium are calculated by solving Eqs. (40) and (41).

### II-4. Dyson's equation for the coherent wavefield

Following the standard procedure of Feynman's diagram technique and the invoking the First-Order-Smoothing approximation (FOSA), as detailed in [13], we obtain the renormalized Dyson's equation for the coherent wave fields

$$\langle \Phi(\mathbf{x}'' - \mathbf{x}') \rangle = \Psi^0(\mathbf{x}'' - \mathbf{x}') + \iiint_{V(\mathbf{y})} P.S.\Gamma(\mathbf{x}'' - \mathbf{y}) \iiint_{V(\mathbf{x})} P.S.\Gamma(\mathbf{y} - \mathbf{x}) \langle \Xi(\mathbf{y})\Xi(\mathbf{x}) \rangle \langle \Phi(\mathbf{x} - \mathbf{x}') \rangle d^3\mathbf{x} . \quad (42)$$

For statistically homogeneous media, the two-point correlation function is dependent on the difference between the two points only, so the Dyson equation can be simplified as

$$\langle \Phi(\mathbf{x}'' - \mathbf{x}') \rangle = \Psi^0(\mathbf{x}'' - \mathbf{x}') + \iiint_{V(\mathbf{y})} P.S.\Gamma(\mathbf{x}'' - \mathbf{y}) \iiint_{V(\mathbf{x})} P.S.\Gamma(\mathbf{y} - \mathbf{x}) \langle \Xi\Xi \rangle P(\mathbf{y} - \mathbf{x}) \langle \Phi(\mathbf{x} - \mathbf{x}') \rangle d^3\mathbf{x} . \quad (43)$$

Applying Fourier transform to the Dyson equation, we get

$$\langle \tilde{\Phi}(\mathbf{k}) \rangle = \tilde{\Psi}^0(\mathbf{k}) + P.S.\tilde{\Gamma}(\mathbf{k}) \langle \Xi\Xi \rangle \left[ \frac{1}{8\pi^3} \iiint_{V(\mathbf{x})} P.S.\tilde{\Gamma}(\mathbf{s})\tilde{P}(\mathbf{k} - \mathbf{s}) d^3\mathbf{s} \right] \langle \tilde{\Phi}(\mathbf{k}) \rangle . \quad (44)$$

Multiplying both sides of this equation by $\left[ P.S.\tilde{\Gamma}(\mathbf{k}) \right]^{-1}$, we get

$$\left[ P.S.\tilde{\Gamma}(\mathbf{k}) \right]^{-1} \langle \tilde{\Phi}(\mathbf{k}) \rangle = \left[ P.S.\tilde{\Gamma}(\mathbf{k}) \right]^{-1} \tilde{\Psi}^0(\mathbf{k}) + \langle \Xi\Xi \rangle \left[ \frac{1}{8\pi^3} \iiint_{V(\mathbf{x})} P.S.\tilde{\Gamma}(\mathbf{s})\tilde{P}(\mathbf{k} - \mathbf{s}) d^3\mathbf{s} \right] \langle \tilde{\Phi}(\mathbf{k}) \rangle. \quad (45)$$

By collecting the coefficient of $\langle \tilde{\Phi}(\mathbf{k}) \rangle$, we obtain an equation for the coherent wavefield in the frequency-wavenumber domain

$$\left\{ \left[ P.S.\tilde{\Gamma}(\mathbf{k}) \right]^{-1} - \langle \Xi\Xi \rangle \left[ \frac{1}{8\pi^3} \iiint_{V(\mathbf{x})} P.S.\tilde{\Gamma}(\mathbf{s})\tilde{P}(\mathbf{k} - \mathbf{s}) d^3\mathbf{s} \right] \right\} \langle \tilde{\Phi}(\mathbf{k}) \rangle = \left[ P.S.\tilde{\Gamma}(\mathbf{k}) \right]^{-1} \tilde{\Psi}^0(\mathbf{k}), \quad (46)$$

The dispersion equation is obtained by setting the coefficient of $\langle \tilde{\Phi}(\mathbf{k}) \rangle$ to zero

$$\det \left\{ \left[ P.S.\tilde{\Gamma}(\mathbf{k}) \right]^{-1} - \langle \Xi\Xi \rangle \left[ \frac{1}{8\pi^3} \iiint_{V(\mathbf{x})} P.S.\tilde{\Gamma}(\mathbf{s})\tilde{P}(\mathbf{k} - \mathbf{s}) d^3\mathbf{s} \right] \right\} = 0 . \quad (47)$$

### II-5. Dispersion equation

For coherent waves propagating along a general direction, the quasi-longitudinal component is coupled to the quasi-transverse components, so the dispersion equation has a rather complicated form. However, different wave modes are decoupled when the wave propagates along the major axes. Consequently, the dispersion equations for each mode are decoupled. For example, we consider coherent waves propagating along the $x_3$ axis, the coefficient matrix in Eq. (46) is simplified as

$$\left[P.S.\tilde{\Gamma}(\mathbf{k})\right]^{-1} - \langle \Xi\Xi \rangle \left[\frac{1}{8\pi^3}\iiint_{V(\mathbf{x})} P.S.\tilde{\Gamma}(\mathbf{s})\tilde{P}(\mathbf{k}-\mathbf{s})d^3\mathbf{s}\right] = \begin{bmatrix} M_{11} & 0 & 0 & 0 & 0 & 0 & 0 & M_{18} & 0 \\ 0 & M_{22} & 0 & 0 & 0 & 0 & M_{27} & 0 & 0 \\ 0 & 0 & M_{33} & M_{34} & M_{35} & M_{36} & 0 & 0 & 0 \\ 0 & 0 & M_{34} & M_{44} & M_{45} & M_{46} & 0 & 0 & 0 \\ 0 & 0 & M_{35} & M_{45} & M_{55} & M_{56} & 0 & 0 & 0 \\ 0 & 0 & M_{36} & M_{46} & M_{56} & M_{66} & 0 & 0 & 0 \\ 0 & M_{27} & 0 & 0 & 0 & 0 & M_{77} & 0 & 0 \\ M_{18} & 0 & 0 & 0 & 0 & 0 & 0 & M_{88} & 0 \\ 0 & 0 & 0 & 0 & 0 & 0 & 0 & 0 & M_{99} \end{bmatrix}, \qquad (48)$$

where

$$M_{11} = c_{55}k^2 - \rho\omega^2 - K_{55}k^2 - (\rho_1-\rho_2)^2 f_1 f_2 \omega^4 \Sigma_{11}, \quad M_{22} = c_{44}k^2 - \rho\omega^2 - K_{44}k^2 - (\rho_1-\rho_2)^2 f_1 f_2 \omega^4 \Sigma_{22}, \qquad (49\text{a})$$

$$M_{18} = -iK_{55}k - (\Xi_{55}^{(1)}-\Xi_{55}^{(2)})(\rho_1-\rho_2)f_1 f_2 \omega^2 \Sigma_{18}, \quad M_{27} = -iK_{44}k - (\Xi_{44}^{(1)}-\Xi_{44}^{(2)})(\rho_1-\rho_2)f_1 f_2 \omega^2 \Sigma_{27}, \qquad (49\text{b})$$

$$M_{33} = c_{33}k^2 - \rho\omega^2 - K_{33}k^2 - (\rho_1-\rho_2)^2 f_1 f_2 \omega^4 \Sigma_{33}, \qquad (49\text{c})$$

$$M_{34} = -iK_{13}k - (\rho_1-\rho_2)(\Xi_{11}^{(1)}-\Xi_{11}^{(2)})f_1 f_2 \omega^2 \Sigma_{34} - (\rho_1-\rho_2)(\Xi_{12}^{(1)}-\Xi_{12}^{(2)})f_1 f_2 \omega^2 \Sigma_{35} - (\rho_1-\rho_2)(\Xi_{13}^{(1)}-\Xi_{13}^{(2)})f_1 f_2 \omega^2 \Sigma_{36}, \qquad (49\text{d})$$

$$M_{35} = -iK_{23}k - (\rho_1-\rho_2)(\Xi_{12}^{(1)}-\Xi_{12}^{(2)})f_1 f_2 \omega^2 \Sigma_{34} - (\rho_1-\rho_2)(\Xi_{22}^{(1)}-\Xi_{22}^{(2)})f_1 f_2 \omega^2 \Sigma_{35} - (\rho_1-\rho_2)(\Xi_{23}^{(1)}-\Xi_{23}^{(2)})f_1 f_2 \omega^2 \Sigma_{36}, \qquad (49\text{e})$$

$$M_{36} = -iK_{33}k - (\rho_1-\rho_2)(\Xi_{13}^{(1)}-\Xi_{13}^{(2)})f_1 f_2 \omega^2 \Sigma_{34} - (\rho_1-\rho_2)(\Xi_{23}^{(1)}-\Xi_{23}^{(2)})f_1 f_2 \omega^2 \Sigma_{35} - (\rho_1-\rho_2)(\Xi_{33}^{(1)}-\Xi_{33}^{(2)})f_1 f_2 \omega^2 \Sigma_{36}, \qquad (49\text{f})$$

$$\begin{aligned}M_{44} &= K_{11} - (\Xi_{11}^{(1)}-\Xi_{11}^{(2)})^2 f_1 f_2 \Sigma_{44} - (\Xi_{12}^{(1)}-\Xi_{12}^{(2)})^2 f_1 f_2 \Sigma_{55} - (\Xi_{13}^{(1)}-\Xi_{13}^{(2)})^2 f_1 f_2 \Sigma_{66} \\ &\quad - 2(\Xi_{11}^{(1)}-\Xi_{11}^{(2)})(\Xi_{12}^{(1)}-\Xi_{12}^{(2)})f_1 f_2 \Sigma_{45} - 2(\Xi_{11}^{(1)}-\Xi_{11}^{(2)})(\Xi_{13}^{(1)}-\Xi_{13}^{(2)})f_1 f_2 \Sigma_{46} - 2(\Xi_{12}^{(1)}-\Xi_{12}^{(2)})(\Xi_{13}^{(1)}-\Xi_{13}^{(2)})f_1 f_2 \Sigma_{56},\end{aligned} \qquad (49\text{g})$$

$$\begin{aligned}M_{55} &= K_{22} - (\Xi_{12}^{(1)}-\Xi_{12}^{(2)})^2 f_1 f_2 \Sigma_{44} - (\Xi_{22}^{(1)}-\Xi_{22}^{(2)})^2 f_1 f_2 \Sigma_{55} - (\Xi_{23}^{(1)}-\Xi_{23}^{(2)})^2 f_1 f_2 \Sigma_{66} \\ &\quad - 2(\Xi_{12}^{(1)}-\Xi_{12}^{(2)})(\Xi_{22}^{(1)}-\Xi_{22}^{(2)})f_1 f_2 \Sigma_{45} - 2(\Xi_{12}^{(1)}-\Xi_{12}^{(2)})(\Xi_{23}^{(1)}-\Xi_{23}^{(2)})f_1 f_2 \Sigma_{46} - 2(\Xi_{22}^{(1)}-\Xi_{22}^{(2)})(\Xi_{23}^{(1)}-\Xi_{23}^{(2)})f_1 f_2 \Sigma_{56},\end{aligned} \qquad (49\text{h})$$

$$\begin{aligned}M_{66} &= K_{33} - (\Xi_{13}^{(1)}-\Xi_{13}^{(2)})^2 f_1 f_2 \Sigma_{44} - (\Xi_{23}^{(1)}-\Xi_{23}^{(2)})^2 f_1 f_2 \Sigma_{55} - (\Xi_{33}^{(1)}-\Xi_{33}^{(2)})^2 f_1 f_2 \Sigma_{66} \\ &\quad - 2(\Xi_{13}^{(1)}-\Xi_{13}^{(2)})(\Xi_{23}^{(1)}-\Xi_{23}^{(2)})f_1 f_2 \Sigma_{45} - 2(\Xi_{13}^{(1)}-\Xi_{13}^{(2)})(\Xi_{33}^{(1)}-\Xi_{33}^{(2)})f_1 f_2 \Sigma_{46} - 2(\Xi_{23}^{(1)}-\Xi_{23}^{(2)})(\Xi_{33}^{(1)}-\Xi_{33}^{(2)})f_1 f_2 \Sigma_{56},\end{aligned} \qquad (49\text{i})$$

$$\begin{aligned}M_{45} &= K_{12} - (\Xi_{11}^{(1)}-\Xi_{11}^{(2)})(\Xi_{12}^{(1)}-\Xi_{12}^{(2)})f_1 f_2 \Sigma_{44} - (\Xi_{12}^{(1)}-\Xi_{12}^{(2)})(\Xi_{22}^{(1)}-\Xi_{22}^{(2)})f_1 f_2 \Sigma_{55} - (\Xi_{13}^{(1)}-\Xi_{13}^{(2)})(\Xi_{23}^{(1)}-\Xi_{23}^{(2)})f_1 f_2 \Sigma_{66} \\ &\quad - [(\Xi_{11}^{(1)}-\Xi_{11}^{(2)})(\Xi_{22}^{(1)}-\Xi_{22}^{(2)}) + (\Xi_{12}^{(1)}-\Xi_{12}^{(2)})^2]f_1 f_2 \Sigma_{45} - [(\Xi_{11}^{(1)}-\Xi_{11}^{(2)})(\Xi_{23}^{(1)}-\Xi_{23}^{(2)}) + (\Xi_{12}^{(1)}-\Xi_{12}^{(2)})(\Xi_{13}^{(1)}-\Xi_{13}^{(2)})]f_1 f_2 \Sigma_{46} \\ &\quad - [(\Xi_{12}^{(1)}-\Xi_{12}^{(2)})(\Xi_{23}^{(1)}-\Xi_{23}^{(2)}) + (\Xi_{13}^{(1)}-\Xi_{13}^{(2)})(\Xi_{22}^{(1)}-\Xi_{22}^{(2)})]f_1 f_2 \Sigma_{56},\end{aligned} \qquad (49\text{j})$$

$$\begin{aligned}M_{46} &= K_{13} - (\Xi_{11}^{(1)}-\Xi_{11}^{(2)})(\Xi_{13}^{(1)}-\Xi_{13}^{(2)})f_1 f_2 \Sigma_{44} - (\Xi_{12}^{(1)}-\Xi_{12}^{(2)})(\Xi_{23}^{(1)}-\Xi_{23}^{(2)})f_1 f_2 \Sigma_{55} - (\Xi_{13}^{(1)}-\Xi_{13}^{(2)})(\Xi_{33}^{(1)}-\Xi_{33}^{(2)})f_1 f_2 \Sigma_{66} \\ &\quad - [(\Xi_{11}^{(1)}-\Xi_{11}^{(2)})(\Xi_{23}^{(1)}-\Xi_{23}^{(2)}) + (\Xi_{12}^{(1)}-\Xi_{12}^{(2)})(\Xi_{13}^{(1)}-\Xi_{13}^{(2)})]f_1 f_2 \Sigma_{45} - [(\Xi_{11}^{(1)}-\Xi_{11}^{(2)})(\Xi_{33}^{(1)}-\Xi_{33}^{(2)}) + (\Xi_{13}^{(1)}-\Xi_{13}^{(2)})^2]f_1 f_2 \Sigma_{46} \\ &\quad - [(\Xi_{12}^{(1)}-\Xi_{12}^{(2)})(\Xi_{33}^{(1)}-\Xi_{33}^{(2)}) + (\Xi_{13}^{(1)}-\Xi_{13}^{(2)})(\Xi_{23}^{(1)}-\Xi_{23}^{(2)})]f_1 f_2 \Sigma_{56},\end{aligned} \qquad (49\text{k})$$

$$\begin{aligned}M_{56} &= K_{23} - (\Xi_{12}^{(1)}-\Xi_{12}^{(2)})(\Xi_{13}^{(1)}-\Xi_{13}^{(2)})f_1 f_2 \Sigma_{44} - (\Xi_{22}^{(1)}-\Xi_{22}^{(2)})(\Xi_{23}^{(1)}-\Xi_{23}^{(2)})f_1 f_2 \Sigma_{55} - (\Xi_{23}^{(1)}-\Xi_{23}^{(2)})(\Xi_{33}^{(1)}-\Xi_{33}^{(2)})f_1 f_2 \Sigma_{66} \\ &\quad - [(\Xi_{12}^{(1)}-\Xi_{12}^{(2)})(\Xi_{23}^{(1)}-\Xi_{23}^{(2)}) + (\Xi_{22}^{(1)}-\Xi_{22}^{(2)})(\Xi_{13}^{(1)}-\Xi_{13}^{(2)})]f_1 f_2 \Sigma_{45} - [(\Xi_{12}^{(1)}-\Xi_{12}^{(2)})(\Xi_{33}^{(1)}-\Xi_{33}^{(2)}) + (\Xi_{13}^{(1)}-\Xi_{13}^{(2)})(\Xi_{23}^{(1)}-\Xi_{23}^{(2)})]f_1 f_2 \Sigma_{46} \\ &\quad - [(\Xi_{22}^{(1)}-\Xi_{22}^{(2)})(\Xi_{33}^{(1)}-\Xi_{33}^{(2)}) + (\Xi_{23}^{(1)}-\Xi_{23}^{(2)})^2]f_1 f_2 \Sigma_{56},\end{aligned} \qquad (49\text{l})$$

$$M_{77} = K_{44} - (\Xi_{44}^{(1)}-\Xi_{44}^{(2)})^2 f_1 f_2 \Sigma_{77}, \quad M_{88} = K_{55} - (\Xi_{55}^{(1)}-\Xi_{55}^{(2)})^2 f_1 f_2 \Sigma_{88}, \quad M_{99} = K_{66} - (\Xi_{66}^{(1)}-\Xi_{66}^{(2)})^2 f_1 f_2 \Sigma_{99}, \qquad (49\text{m})$$

and

$$K_{11} = \frac{S_{23}^2 - S_{22}S_{33}}{\Delta_K}, \quad K_{22} = \frac{S_{13}^2 - S_{11}S_{33}}{\Delta_K}, \quad K_{33} = \frac{S_{12}^2 - S_{11}S_{22}}{\Delta_K}, \quad K_{12} = \frac{S_{12}S_{33} - S_{13}S_{23}}{\Delta_K}, \quad K_{13} = \frac{S_{13}S_{22} - S_{12}S_{23}}{\Delta_K}, \quad K_{23} = \frac{S_{23}S_{11} - S_{12}S_{13}}{\Delta_K},$$

$$K_{44} = \frac{1}{4S_{44}}, \quad K_{55} = \frac{1}{4S_{55}}, \quad K_{66} = \frac{1}{4S_{66}}, \quad \Delta_K = S_{11}S_{23}^2 + S_{22}S_{13}^2 + S_{33}S_{12}^2 - 2S_{12}S_{13}S_{23} - S_{11}S_{22}S_{33}, \qquad (50)$$

$$\Sigma_{11} = \frac{1}{8\pi^3}\iiint_{V(\mathbf{s})} \tilde{G}_{11}(\mathbf{s})\tilde{P}(\mathbf{k}-\mathbf{s})d^3\mathbf{s}, \quad \Sigma_{22} = \frac{1}{8\pi^3}\iiint_{V(\mathbf{s})} \tilde{G}_{22}(\mathbf{s})\tilde{P}(\mathbf{k}-\mathbf{s})d^3\mathbf{s}, \quad \Sigma_{18} = \frac{1}{8\pi^3}\iiint_{V(\mathbf{s})} [is_3\tilde{G}_{11}(\mathbf{s}) + is_1\tilde{G}_{13}(\mathbf{s})]\tilde{P}(\mathbf{k}-\mathbf{s})d^3\mathbf{s}, \qquad (51\text{a})$$

$$\Sigma_{27} = \frac{1}{8\pi^3} \iiint_{V(s)} [is_3\tilde{G}_{22}(\mathbf{s}) + is_2\tilde{G}_{23}(\mathbf{s})]\tilde{P}(\mathbf{k}-\mathbf{s})d^3\mathbf{s}, \quad \Sigma_{33} = \frac{1}{8\pi^3}\iiint_{V(s)} \tilde{G}_{33}(\mathbf{s})\tilde{P}(\mathbf{k}-\mathbf{s})d^3\mathbf{s}, \tag{51b}$$

$$\Sigma_{34} = \frac{1}{8\pi^3}\iiint_{V(s)} is_1\tilde{G}_{13}(\mathbf{s})\tilde{P}(\mathbf{k}-\mathbf{s})d^3\mathbf{s}, \quad \Sigma_{35} = \frac{1}{8\pi^3}\iiint_{V(s)} is_2\tilde{G}_{23}(\mathbf{s})\tilde{P}(\mathbf{k}-\mathbf{s})d^3\mathbf{s}, \quad \Sigma_{36} = \frac{1}{8\pi^3}\iiint_{V(s)} is_3\tilde{G}_{33}(\mathbf{s})\tilde{P}(\mathbf{k}-\mathbf{s})d^3\mathbf{s}, \tag{51c}$$

$$\Sigma_{44} = S_{11} - \frac{1}{8\pi^3}\iiint_{V(s)} s_1^2\tilde{G}_{11}(\mathbf{s})\tilde{P}(\mathbf{k}-\mathbf{s})d^3\mathbf{s}, \quad \Sigma_{55} = S_{22} - \frac{1}{8\pi^3}\iiint_{V(s)} s_2^2\tilde{G}_{22}(\mathbf{s})\tilde{P}(\mathbf{k}-\mathbf{s})d^3\mathbf{s}, \tag{51d}$$

$$\Sigma_{66} = S_{33} - \frac{1}{8\pi^3}\iiint_{V(s)} s_3^2\tilde{G}_{33}(\mathbf{s})\tilde{P}(\mathbf{k}-\mathbf{s})d^3\mathbf{s}, \quad \Sigma_{45} = S_{12} - \frac{1}{8\pi^3}\iiint_{V(s)} s_1 s_2\tilde{G}_{12}(\mathbf{s})\tilde{P}(\mathbf{k}-\mathbf{s})d^3\mathbf{s}, \tag{51e}$$

$$\Sigma_{46} = S_{13} - \frac{1}{8\pi^3}\iiint_{V(s)} s_1 s_3\tilde{G}_{13}(\mathbf{s})\tilde{P}(\mathbf{k}-\mathbf{s})d^3\mathbf{s}, \quad \Sigma_{56} = S_{23} - \frac{1}{8\pi^3}\iiint_{V(s)} s_2 s_3\tilde{G}_{13}(\mathbf{s})\tilde{P}(\mathbf{k}-\mathbf{s})d^3\mathbf{s}, \tag{51f}$$

$$\Sigma_{77} = 4S_{44} - \frac{1}{8\pi^3}\iiint_{V(s)}\left[s_3^2\tilde{G}_{22}(\mathbf{s}) + s_2^2\tilde{G}_{33}(\mathbf{s}) + 2s_2 s_3\tilde{G}_{23}(\mathbf{s})\right]\tilde{P}(\mathbf{k}-\mathbf{s})d^3\mathbf{s}, \tag{51g}$$

$$\Sigma_{88} = 4S_{55} - \frac{1}{8\pi^3}\iiint_{V(s)}\left[s_3^2\tilde{G}_{11}(\mathbf{s}) + s_1^2\tilde{G}_{33}(\mathbf{s}) + 2s_1 s_3\tilde{G}_{13}(\mathbf{s})\right]\tilde{P}(\mathbf{k}-\mathbf{s})d^3\mathbf{s}, \tag{51h}$$

$$\Sigma_{99} = 4S_{66} - \frac{1}{8\pi^3}\iiint_{V(s)}\left[s_2^2\tilde{G}_{11}(\mathbf{s}) + s_1^2\tilde{G}_{22}(\mathbf{s}) + 2s_1 s_2\tilde{G}_{12}(\mathbf{s})\right]\tilde{P}(\mathbf{k}-\mathbf{s})d^3\mathbf{s}. \tag{51i}$$

In the derivation of Eq. (51), the relation

$$\frac{1}{8\pi^3}\iiint_{V(s)} S_{11}\tilde{P}(\mathbf{k}-\mathbf{s})d^3\mathbf{s} = \frac{1}{8\pi^3}\iiint_{V(s)} S_{11}\cdot 1\cdot \tilde{P}(\mathbf{k}-\mathbf{s})d^3\mathbf{s} = \iiint_{V(x)} S_{11}\cdot\delta(\mathbf{x})\cdot P(\mathbf{x})e^{-i\mathbf{k}\cdot\mathbf{x}}d^3\mathbf{x} = S_{11}P(\mathbf{0})e^{-i\mathbf{k}\cdot\mathbf{0}} = S_{11} \tag{52}$$

is used to simplify the final results.

The dispersion equation (47) can be factorized into a product of three factors, each corresponds to a specific wave mode. The dispersion equation of longitudinal waves is

$$\begin{aligned}&M_{33}(M_{44}M_{56}^2 + M_{55}M_{46}^2 + M_{66}M_{45}^2 - 2M_{45}M_{46}M_{56} - M_{44}M_{55}M_{66})\\&+ M_{34}^2(M_{55}M_{66} - M_{56}^2) + M_{35}^2(M_{44}M_{66} - M_{46}^2) + M_{36}^2(M_{44}M_{55} - M_{45}^2)\\&+ 2M_{34}M_{35}(M_{46}M_{56} - M_{45}M_{66}) + 2M_{34}M_{36}(M_{45}M_{56} - M_{46}M_{55}) + 2M_{35}M_{36}(M_{45}M_{46} - M_{56}M_{44}) = 0,\end{aligned} \tag{53}$$

The dispersion equation of transverse waves polarized along the $x_1$ axis ($T_{31}$ wave) is

$$M_{11}M_{88} - M_{18}^2 = 0, \tag{54}$$

The dispersion equation of transverse waves polarized along the $x_2$ axis ($T_{32}$ wave) is

$$M_{22}M_{77} - M_{27}^2 = 0, \tag{55}$$

The solution of the dispersion equations is a complex wavenumber, the real part corresponds to the velocity, and the imaginary part represents the attenuation. In the following discussion we introduce the dimensionless fractional velocity variation $\delta V = (V - V_0)/V_0$, the dimensionless attenuation coefficient $\alpha d$, and the inverse Q-factor $Q^{-1}$, where $V_0$ is the velocity of the reference medium, $d$ is the characteristic diameter of the inhomogeneities [13]. Now we are in a position to discuss the numerical solution of the dispersion equations. Consider longitudinal waves propagating along the $x_3$ axis, i.e., $\mathbf{k} = k\mathbf{e}_3$. We can define the following dimensionless variables

$$K_{0L} = k_L a_z, \quad k_L = \frac{\omega}{V_L}, \quad V_L = \sqrt{\frac{c_{33}}{\rho}}, \tag{56}$$

Next, we can introduce the following nonorthogonal ellipsoidal coordinate system

$$s_x = \frac{1}{a_x}K_{0L}S\sin\theta\cos\varphi, \quad s_y = \frac{1}{a_y}K_{0L}S\sin\theta\sin\varphi, \quad s_z = \frac{1}{a_z}K_{0L}S\cos\theta. \tag{57}$$

In the new coordinate system, the Christoffel tensor takes the form

$$\Gamma_{11} = \frac{c_{11}}{a_x^2}K_{0L}^2 S^2\sin^2\theta\cos^2\varphi + \frac{c_{66}}{a_y^2}K_{0L}^2 S^2\sin^2\theta\sin^2\varphi + \frac{c_{55}}{a_z^2}K_{0L}^2 S^2\cos^2\theta, \tag{58a}$$

$$\Gamma_{22} = \frac{c_{66}}{a_x^2} K_{0L}^2 S^2 \sin^2\theta \cos^2\varphi + \frac{c_{22}}{a_y^2} K_{0L}^2 S^2 \sin^2\theta \sin^2\varphi + \frac{c_{44}}{a_z^2} K_{0L}^2 S^2 \cos^2\theta, \tag{58b}$$

$$\Gamma_{33} = \frac{c_{55}}{a_x^2} K_{0L}^2 S^2 \sin^2\theta \cos^2\varphi + \frac{c_{44}}{a_y^2} K_{0L}^2 S^2 \sin^2\theta \sin^2\varphi + \frac{c_{33}}{a_z^2} K_{0L}^2 S^2 \cos^2\theta, \tag{58c}$$

$$\Gamma_{12} = \frac{c_{12}+c_{66}}{a_x a_y} K_{0L}^2 S^2 \sin^2\theta \sin\varphi\cos\varphi,\ \Gamma_{13} = \frac{c_{13}+c_{55}}{a_x a_z} K_{0L}^2 S^2 \sin\theta\cos\theta\cos\varphi,\ \Gamma_{23} = \frac{c_{23}+c_{44}}{a_y a_z} K_{0L}^2 S^2 \sin\theta\sin\varphi\cos\theta. \tag{58d}$$

We are interested in the dynamic behavior of the wave propagation, so we can choose $K_{0L} \neq 0$. The homogeneous Green's function can be rewritten as

$$\tilde{G}_{ij}^0(\mathbf{s}) = \frac{a_z^2}{c_{33}} \frac{1}{K_{0L}^2} \sum_{n=1}^{3} \frac{1}{\lambda_n} \frac{q_i^{(n)} q_j^{(n)}}{S^2 - \frac{1}{\lambda_n}}, \tag{59}$$

Similarly, for transverse wave T$_{31}$, we can introduce the dimensionless quantities

$$K_{0T} = k_T a_z\ ,\ k_T = \frac{\omega}{V_T}\ ,\ V_T = \sqrt{\frac{c_{11}}{\rho}}\ , \tag{60}$$

and the wavenumber $s_i$ and Christoffel tensor can be obtained by replacing $K_{0L}$ in Eqs. (57) and (58) by $K_{0T}$. When the dimensionless wavenumber is introduced, the integrals in Eq. (51) are triple integrals of three independent variables: $S$, $\theta$, and $\varphi$. Different from the integrals appeared in the dispersion equation for isotropic media [13], in which the triple integrals can be split into two parts, one dependent on $S$ only and the other dependent on $\theta$ and $\varphi$, the integrals here cannot be decoupled, so we first need to specify the two angles $\theta$ and $\varphi$, performing eigenspace decomposition of the Green's functions, and calculate the integral with respect to $S$, and then finish the integral over $\theta$ and $\varphi$. As a result, the computation cost of anisotropic media is much higher than that for the isotropic case. We also need to mention that the integral over $S$ is singular at the points $S = 1/\sqrt{\lambda_n}$, so the Cauchy principle value must be adopted. Readers interested in the detailed integral technique are referred to [13].

## III. Numerical results and applications

Multiple scattering theories have found applications in a broad spectrum of distinct disciplines. For example, multiple scattering of electrons and neutrons in solid have found applications in nuclear and semiconductor physics [16]. Multiple scattering of electromagnetic waves is used in characterization of dielectric materials and in satellite remote sensing [10-11, 17]. Multiple scattering of ultrasonic waves is widely used in nondestructive characterization of polycrystal microstructures [6-7, 9, 11]. Nevertheless, our attenuation in this work will be mainly focused on its applications in seismology. Scattering and attenuation of the seismic waves is one of the central research topics in seismology [18-25]. The multiple scattering phenomenon observed in seismograms recorded worldwide and the accompanying long coda waves after the major seismic phases indubitably reveal the highly inhomogeneous nature of the Earth. Aki's pioneering research on coda waves first reveals that small-scale heterogeneities in the Earth's lithosphere play a key role in determining the scattering and attenuation of seismic waves [19]. Although it is widely acknowledged that scattering and intrinsic absorption both contribute to the seismic wave attenuation, the dominant mechanism is still a longstanding problem [23]. Pioneering seismologists, represented by Aki [18-20], Wu [22-25], and Sato [21] all devoted significant efforts to study the scattering, attenuation, and energy transport of seismic waves. Although meaningful progresses have been achieved, it is generally believed that many problems are still open. This research area remains active up to date.

The multiple scattering theory developed in this work establishes a general and powerful framework to study seismic wave propagation in the heterogeneous Earth. First and for most, it gives up the weak scattering approximation and thus, enables quantitative analysis of velocity and attenuation by using strong scattering models of the lithosphere. Second, it properly incorporates the effects of non-equiaxed heterogeneities, which makes statistical characterization of geometric anisotropy of small-scale heterogeneities possible. As pointed out in Sec. I, both sonic log measurements and traveltime tomography reveal that non-equiaxed small-scale heterogeneities exist in some regions of lithosphere. Their aspect ratio can vary from near unity to 7 or even larger. The length scale of these heterogeneities covers several orders of magnitude. However, we assume that the heterogeneities with characteristic size about several hundred meters to tens of kilometers play a key role in determining the seismic scattering and attenuation because the wavelength of typical seismic waves lies in this region. Practical measurements and imaging results shows that the velocity of P waves in the lithosphere varies from 6 to 8 km/s, and the velocity of the S waves lies between 3.5 to 4.5 km/s. The density is related to the velocity by the Birch's law, which states that the density and velocity of Earth medium have a quasilinear relation, and materials with larger density also has larger velocity. It is well known that the density in the lithosphere lies between 2.8 to 3.5 g/cm$^3$. Considering all these facts, we study a medium model which is constituted of two different materials with equal volume fraction, i.e., $f_1=f_2=50\%$. The material properties are shown in Tab. 1. This model can properly simulate the property fluctuation of real rocks in Earth medium.

Table 1   Mechanical properties of the medium model

| | Density: $\rho$ (kg/m$^3$) | $\lambda$ (GPa) | $\mu$ (GPa) | $V_L$ (m/s) | $V_T$ (m/s) |
|---|---|---|---|---|---|
| Phase 1 | 3500 | 100 | 65 | 8106.43 | 4309.46 |
| Phase 2 | 3000 | 40 | 35 | 6055.30 | 3415.65 |

Practical observation of numerous natural earthquakes shows the wave propagation in the lithosphere exhibits roughly isotropic behavior in the horizontal directions, so we consider heterogeneities with the same horizontal characteristic size. Here the coordinate system is chosen such that the y axis coincides with the vertical axis perpendicular to Earth's surface, so we have $a_x=a_z$. To study the effects of geometric anisotropy of the heterogeneities on the wave propagation, we consider heterogeneities with various vertical characteristic size, with the aspect ratio $a_z:a_y$ varying from 1 to 7. The material properties of the reference media are obtained by solving Eqs. (40) and (41). The density of the reference medium is independent of the aspect ratio, which is 3250 kg/m$^3$ for all the cases. However, the elastic stiffness is strongly dependent on the aspect ratio. Table 2 shows the elastic stiffness of the considered medium models. It is seen that the reference medium is transversely isotropic. The tensile moduli along the horizontal axes increases when the aspect ratio becomes larger. Meanwhile, the modulus along the vertical axis is weakened. The shear modulus $C_{55}$ of in-plane ($x_1 x_3$ plane) deformation increases with the aspect ratio, and the shear moduli for the out-of-plane deformation are weakened.

Table 2 Mechanical properties of the reference media, Unit: GPa

| $a_x{:}a_y{:}a_z$ | $C_{11}$ | $C_{22}$ | $C_{33}$ | $C_{12}$ | $C_{13}$ | $C_{23}$ | $C_{44}$ | $C_{55}$ | $C_{66}$ |
|---|---|---|---|---|---|---|---|---|---|
| 1:1:1 | 157.44 | 157.44 | 157.44 | 61.90 | 61.90 | 61.90 | 47.77 | 47.77 | 47.77 |
| 3:1:3 | 160.43 | 152.77 | 160.43 | 61.06 | 63.04 | 61.06 | 47.06 | 48.70 | 47.06 |
| 5:1:5 | 161.69 | 151.35 | 161.69 | 60.61 | 63.52 | 60.61 | 46.64 | 49.09 | 46.64 |
| 7:1:7 | 162.43 | 150.68 | 162.40 | 60.37 | 63.81 | 60.36 | 46.40 | 49.30 | 46.38 |

Without loss of generality, we consider the coherent waves propagating along the $x_3$ axis. This theoretical setup exactly reproduces the observation configuration of real earthquakes. The reference velocities of the medium models are shown in Tab. 3. Compared with the equiaxed medium, change in the velocity reaches up to 100 m/s solely due to the geometric anisotropy.

Table 3 Wave velocities of the reference media, Unit: m/s

| $a_x{:}a_y{:}a_z$ | $V_{L3}$ | $V_{T31}$ | $V_{T32}$ |
|---|---|---|---|
| 1:1:1 | 6960.73 | 3834.46 | 3834.46 |
| 3:1:3 | 7025.89 | 3871.00 | 3805.26 |
| 5:1:5 | 7053.42 | 3886.47 | 3788.24 |
| 7:1:7 | 7068.89 | 3894.77 | 3778.48 |

To show the effects of aspect ratio on the dispersion and attenuation behavior of the scattered waves, we first replot the dispersion and attenuation curves for the equiaxed medium model for reference, which was denoted by Medium XII in [13], see Fig. 2. We need to mention that the dispersion equation for the anisotropic media cannot degenerate into the isotropic case since ill-defined factors appear in the dispersion equations.

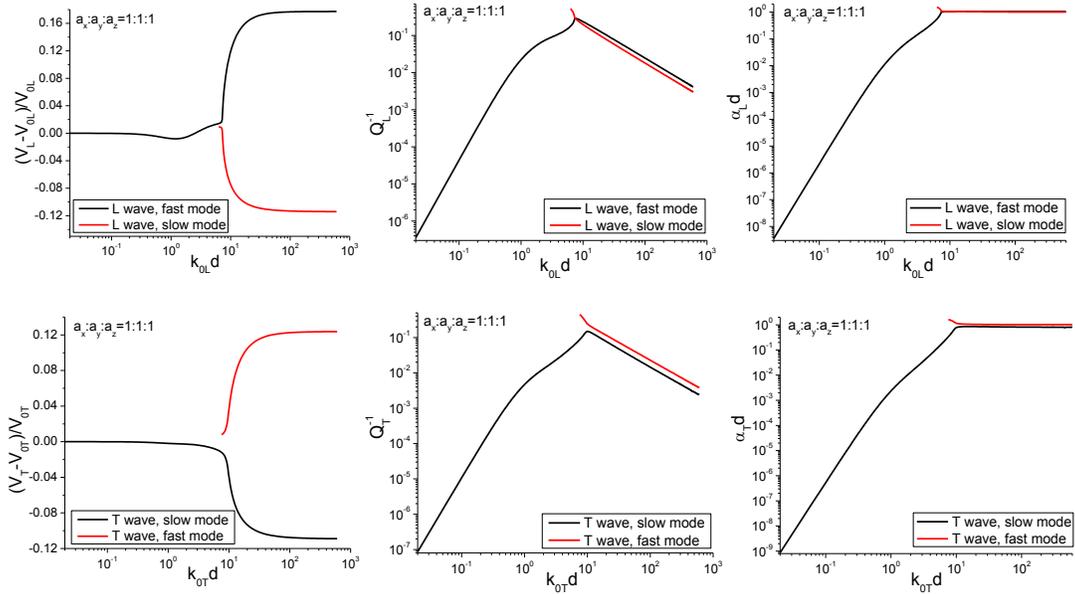

Figure 2 Velocity and attenuation of the medium model when $a_x{:}a_y{:}a_z$=1:1:1

The dispersion and attenuation curves for non-equiaxed heterogeneous media with different aspect ratios are shown in Figs. 3-5. From the numerical results we can observe that the coherent wave velocity and attenuation share some common features. It is seen that there is only one propagation mode in the frequency range 0<$k_{0L}d$, $k_{0T}d$<5, the dispersion of this mode is very small, and its velocity varies near the quasi-static limit. Both the Q-factor and the attenuation coefficient increase with the frequency following a power law. This is typical for waves in the Rayleigh scattering regime. In the frequency band 5<$k_{0L}d$, $k_{0T}d$<30, which is also known as the stochastic scattering region, the dispersion of the single wave mode increases dramatically, and the single wave mode splits into two. The velocity of the fast mode approaches the upper limit of the component materials, i.e., Phase 1, and the velocity of the slow mode approaches the lower bound of the component material, Phase 2. The Q-factors assume their maximum value at around $k_{0L}d$, $k_{0T}d=10$, at which the wave length is equal to the characteristic size of the inhomogeneities. This indicates the occurrence of the so-called resonant scattering. In the high frequency region, $k_{0L}d$, $k_{0T}d<30$, the dispersion of the two modes becomes very small again, and the Q-factors decrease with frequency following an inverse power law. The attenuation coefficient stabilized near unity. This is typical for waves in the geometric scattering regime. Meanwhile, we find that variations in the heterogeneity aspect ratio result in obvious differences among the dispersion and attenuation curves. The dispersion of non-equiaxed heterogeneous media in the stochastic regime is smaller than the medium with equiaxed inhomogeneities. The maximum value of the Q-factors decreases from 0.2 to 0.05. Meanwhile, the sharp peak observed in the Q-factor of the longitudinal waves in the equiaxed medium disappears, the Q-factors for longitudinal waves of all the non-equiaxed medium models exhibit a flat plateau near the maximum value. The Q-factors of the transverse waves for non-equiaxed media still have a sharp peak. All these characteristics tell us that the wave components in the stochastic transition regime in non-equiaxed medium may be much easier to observe than those in equiaxed medium. In the geometric regime, the magnitude of the Q-factors of both the longitudinal and the slow transverse waves decreases dramatically compared to that of the equiaxed media. This can also be found from the attenuation curves, where the attenuation coefficient of the longitudinal waves is decreased from 1 to 0.3 or even smaller. The attenuation coefficient of the slow transverse mode is decreased from 1 to 0.3, while that of the fast mode still lies above 1.

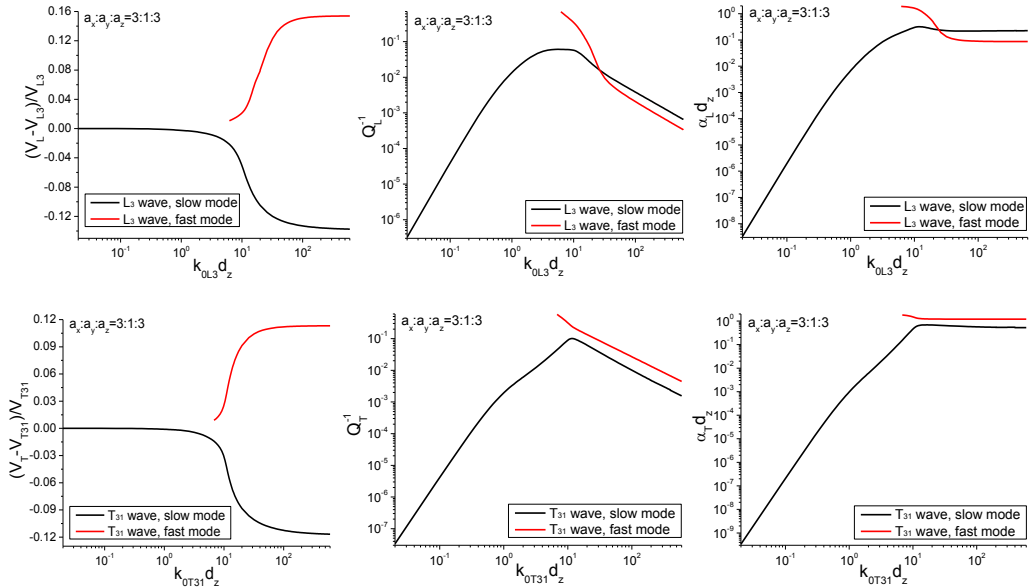

Figure 3 Velocity and attenuation of the medium model when $a_x:a_y:a_z=3:1:3$

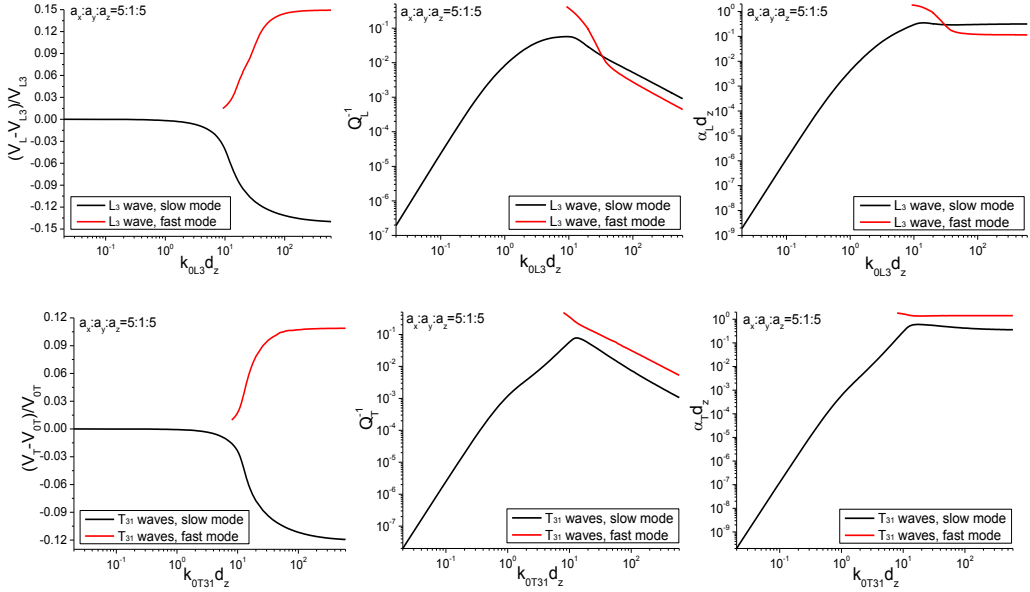

Figure 4 Velocity and attenuation of the medium model when $a_x:a_y:a_z =5:1:5$

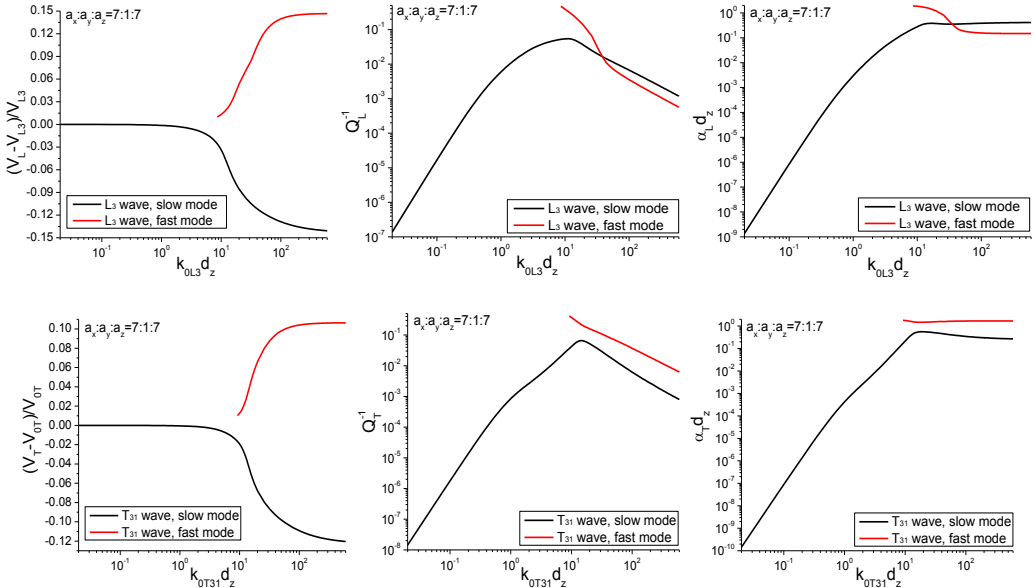

Figure 5 Velocity and attenuation of the medium model when $a_x:a_y:a_z =7:1:7$

The longitudinal-to-transverse Q-factor ratios for the considered medium models are plotted in Fig. 6. The numerical results reveal the following features of the Q-factor ratios: 1) The quasi-static limit of the Q-factor ratio for the equiaxed medium lies near 0.5, but the ratio for non-equiaxed media lies near 1.5; 2) The Q-factor ratio for all the cases exhibits a peak near $k_{0L}d=1.5$, which tells us that near this frequency the maximum energy transfer from the longitudinal waves to transverse waves occurs; 3) All the Q-factor ratios show a local minimum near $k_{0L}d=7$, at this frequency the attenuation of the transverse waves achieves its maximum; 4) In the geometric regime, the Q-factor ratios of the equiaxed medium approaches four different values, and all of them are larger than unity. However, the Q-factor ratios for non-equiaxed media show a more complicated pattern. The most obvious point is that the ratios of certain mode combinations can be smaller than unity, varying from 0.1 to 1. Another important feature is that when the aspect ratio is very large, for example, $a_z:a_y$=5-7, the Q-factor ratio for the slow longitudinal mode to the slow transverse mode does not stabilize at a constant value, instead it increases with frequency.

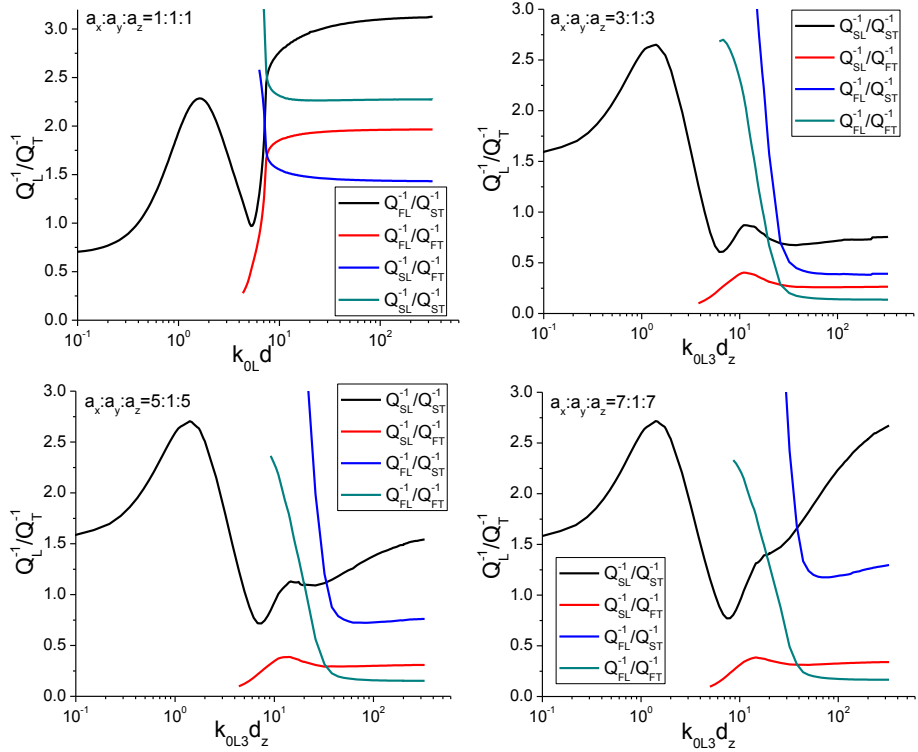

Figure 6 Ratios of the longitudinal to transverse Q-factors

In the above discussion, the frequency band is divided into Rayleigh, stochastic and geometric regimes based on the dimensionless frequency $kd$, which is dependent on the frequency, wave velocity, and characteristic size of heterogeneities. In order to gain an intuitive understanding of the numerical results, we consider a standard model, for which the velocities are: $V_{0L}$=7000 m/s, $V_{0T}$=3800 m/s, the dimensionless critical frequencies are $K_{cr}^L = 10$, $K_{cr}^T = 12$, respectively. The characteristic size, or called the average diameter, of the heterogeneities in the lithosphere may vary from 2 km to 100 km. In Tabs. 4 and 5 we list the scattering regimes in terms of the physical frequency for a series of characteristic sizes. The tables clearly show that the scattering regime is strongly dependent on the characteristic size. For example, for medium with d=2 km, the high frequency regime for longitudinal waves starts from 16.71 Hz, however, for medium with d= 50 km, longitudinal waves at 0.67 Hz is already in the high frequency regime.

Table 4 Scattering frequency band (Hz) of the standard model with various characteristic sizes

| Characteristic size | Critical frequency | | Low frequency | | Intermediate frequency | | High frequency | |
|---|---|---|---|---|---|---|---|---|
| $d_z$ (km) | $f_{cr}^L$ | $f_{cr}^T$ | $K_{0L}$ = [0.01, 1] | $K_{0T}$ = [0.01, 1] | $K_{0L}$ = [1, 30] | $K_{0T}$ = [1, 30] | $K_{0L}$ = [30, 2000] | $K_{0T}$ = [30, 2000] |
| 2 | 5.57 | 3.63 | [5.57 × 10⁻³, 0.56] | [3.02 × 10⁻³, 0.30] | [0.56, 16.71] | [0.30, 9.07] | [16.71, 1114] | [9.07, 604.79] |
| 5 | 2.23 | 1.45 | [2.23 × 10⁻³, 0.22] | [1.21 × 10⁻³, 0.12] | [0.22, 6.68] | [0.12, 3.63] | [6.68, 445.63] | [3.63, 241.92] |
| 10 | 1.11 | 0.73 | [1.11 × 10⁻³, 0.11] | [6.05 × 10⁻⁴, 0.06] | [0.11, 3.34] | [0.06, 1.81] | [3.34, 222.82] | [1.81, 120.96] |
| 30 | 0.37 | 0.24 | [3.71 × 10⁻⁴, 0.04] | [2.02 × 10⁻⁴, 0.02] | [0.04, 1.11] | [0.02, 0.60] | [1.11, 74.27] | [0.60, 40.32] |
| 50 | 0.22 | 0.15 | [2.23 × 10⁻⁴, 0.02] | [1.21 × 10⁻⁴, 0.012] | [0.02, 0.67] | [0.012, 0.36] | [0.67, 44.56] | [0.36, 24.19] |

| 70 | 0.16 | 0.10 | [1.59 × 10⁻⁴, 0.016] | [8.64 × 10⁻⁵, 0.0086] | [0.016, 0.48] | [0.0056, 0.26] | [0.48, 31.83] | [0.26, 17.28] |
| 100 | 0.11 | 0.07 | [1.11 × 10⁻⁴, 0.011] | [6.05 × 10⁻⁵, 0.0051] | [0.011, 0.33] | [0.0051, 0.18] | [0.33, 22.28] | [0.18, 12.10] |

Table 5 Dimensionless frequency v.s. frequency for various characteristic sizes

| $d$ (km) | 2 | | 5 | | 10 | | 30 | | 50 | | 70 | | 100 | |
|---|---|---|---|---|---|---|---|---|---|---|---|---|---|---|
| $f$ (Hz) | $K_{0L}$ | $K_{0T}$ | $K_{0L}$ | $K_{0T}$ | $K_{0L}$ | $K_{0T}$ | $K_{0L}$ | $K_{0T}$ | $K_{0L}$ | $K_{0T}$ | $K_{0L}$ | $K_{0T}$ | $K_{0L}$ | $K_{0T}$ |
| 0.01 | 0.018 | 0.033 | 0.045 | 0.083 | 0.090 | 0.165 | 0.27 | 0.50 | 0.45 | 0.83 | 0.63 | 1.16 | 0.90 | 1.65 |
| 1 | 1.80 | 3.31 | 4.49 | 8.27 | 8.98 | 16.53 | 26.93 | 49.60 | 44.88 | 82.67 | 62.83 | 115.74 | 89.76 | 165.35 |
| 10 | 17.95 | 33.07 | 44.88 | 82.67 | 89.76 | 165.35 | 269.28 | 496.04 | 448.80 | 826.73 | 628.32 | 1157 | 897.60 | 1653 |
| 100 | 179.52 | 330.69 | 448.80 | 826.73 | 897.60 | 1653 | 2692 | 4960 | 4488 | 8267 | 6283 | 11574 | 8975 | 16534 |

The numrical results provide us important insight into the statistical characfteristics of the subsurface heterogeneities. They also help us get a better understanding on the scattering and attenuation behaviors of real seismic waves. Sato et. al. [21] collected the seismic data of major seismic phases in numerious reginonal and global rarthquakes and plotted the Q-factors and their ratios, as shown in Fig. 7. From the observed results we can see both the longitudinal and transverse Q-factors decrease with frequency following an inverse power law, indicating that most of the seismic events lie in the geometric regime. The longitudinal Q-factors exhibit a flat plateau in the frequency range 0.1 Hz< $f$ <1 Hz, while the transverse Q-factors still show one sharp peak. One possible explanation is that the subsurface inhomogeneities are non-equiaxed, e.g., of oblate spheridal geometry. The geometric anisotropy eliminates the sharp peak of the longitudinal Q-factors while still keeping the transverse Q-factors observable. The low frequency components of natural earthquakes are extremely difficult to measure, so the information in this regime are missing. Figure 7(c) shows the longitudinal-to-transverse Q-factor ratios obtained from real earthquakes. Through comparison with the numerical results in Fig. 6, we can establish the following correspondence: 1) Curves 1, 3, 2.1 and 2.2 lie in the frequency band $k_{0L}d < 2$, in which the Q-factor ratio increases from 0.5 to 1.5 or even larger; 2) Curves 5, 18, 14 lie in the frequency band $2<k_{0L}d < 20$, in which the ratios reach the maximum value and decreases to its minimum, and then increase again; 3) Curves 4, 8, 10, 6, 13, 17.1, 17.2 lie in the geometric regime, in which the Q-factor ratios are constant. In particular, we need to mention that the curves 13, 17.1, and 17.2 stabilize near or below unity, indicating the aspect ratios in these areas are larger than unity. Judging based on the numerical results shown in Fig. 6, Curve 16 may also lie in the geometric regime because the $Q_{SL}^{-1}/Q_{ST}^{-1}$ curve for aspect ratio $a_z:a_y=7$ increases with frequency in the geometric regime. Curve 15 shows Q-factor ratio decreasing very fast, this may correspond to the curve $Q_{FL}^{-1}/Q_{ST}^{-1}$ in Fig. 6. From the above analysis we see that the new scattering model is able to explain all the variation tendency of the Q-factors and their ratios. Moreover, the magnitudes of these quantities also show excellent agreement with that predicted by the model. All these facts are strong evidence to show that the dominant mechanism for seismic wave attenuation is scattering, the portion contributed by intrinsic mechanisms is less important. We can further give a rough estimation of the characteristic length scale of the subsurface heterogeneities with the help of Tabs 4 and 5, interested readers are referred to [13].

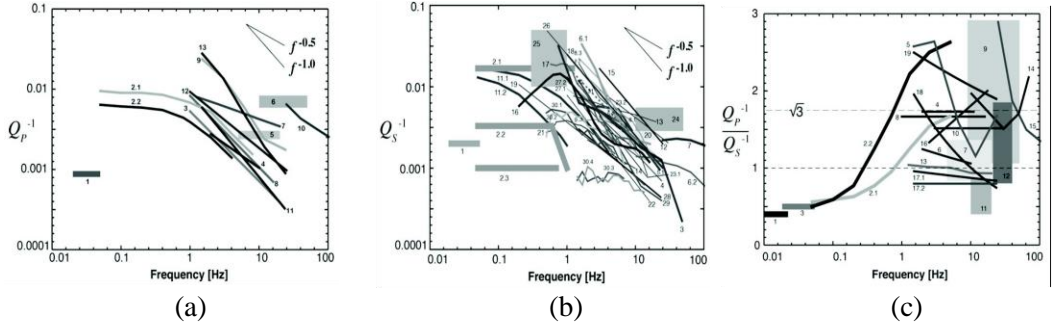

FIG. 7. Q-factors of longitudinal (a) and transverse (b) waves, and their ratios (c) measured in local, regional and global earthquakes [21].

Accurate measurement of dispersions of seismic waves is very challenging since the surface irregularities and near receiver disturbances often destroy the phase information of the major arrivals. Here we will not get involved in the detailed correction and extraction of dispersion curves, instead, we will focus on the most striking phenomenon occurred in the high frequency regime - the splitting of a single wave package into two! Figure 8 shows the traveltime curves recorded for an earthquake of magnitude 8 occurred in 1908, which was obtained by Andrija Mohorovičić [26]. It shows that there is only one arrival recorded from epicentral distance of less than 300 km. But seismic stations with epicentral distance greater than 300 km recorded four arrivals, denoted by Pn, Pg, Sn and Sg, respectively. Since then, similar traveltime curves are recorded from seismic station worldwide for numerous seismic events, see Figs. 9-10. Two additional phases $P^*$ and $S^*$ are observed in Fig. 10(c) with velocity lying between the Pn and Pg or Sn and Sg phases. Mohorovičić proposed an explanation based on the multilayered model of the Earth lithosphere. He thought that there is a discontinuity about 50km underneath the Earth surface, which separate two different media with drastically different properties. The wave velocity of each layer is detected from the velocity of the different seismic phases. This discontinuity is named the Mohorovičić discontinuity. Conrad first discovered the $P^*$ and $S^*$ phases and introduced an additional continuity called the Conrad discontinuity. Although the multi-layered model provides a possible model to explain the seismic phases, it is still full of controversy. First, the physical mechanism for the formation of the discontinuities are not clear. Both phase transition and chemical composition change encounter difficulties in explaining such sharp discontinuities. Second, the attenuation predicted by the multi-layered model is in contradiction with observed data, see Aki [18], vol. 1, pp. 529-530. The attenuation of the head wave propagating along the Moho discontinuity is much larger than that observed in real earthquakes. Furthermore, the observed amplitude of the reflected waves from the "discontinuities" are much smaller than that predicted by the multi-layered model, see Aki [18], vol. 1, pp. 212-214.

The new model gives us a new formation mechanism of the seismic phases. As can be seen from the Q-factor curves, most seismic events lie in the geometric regime. According to the new model, the spectrum of the coherent longitudinal and transverse waves in this frequency regime split into two branches. The high-frequency seismic phases Pn, Pg and Sn and Sg exactly correspond to these wave modes. The velocities of these modes reflect the velocity fluctuation of the heterogeneous lithosphere. When the seismic source contains both low and intermediate frequency components, two additional phases $P^*$ and $S^*$ corresponding to the low frequency branch of the dispersion curves can be observed. The new model gives a consistent explanation to the major seismic phases and their attenuation. Most important of all, it is based on the more realistic model as confirmed by practical seismic imaging. Based on the excellent agreement between the numerical results and the practical measurement, we can conclude that the lithosphere is a highly heterogeneous medium, with the characteristic size of the inhomogeneities varying from several kilometers to tens of kilometers, and the velocity fluctuation can be as large as ± 10%. The aspect ratio of the inhomogeneities can vary from unity to as large as 7. All these statistical features can be used to classify the lithosphere.

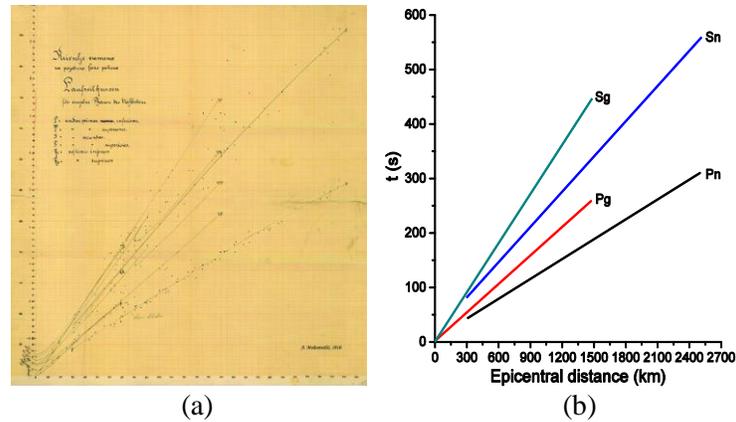

(a) (b)

FIG. 8. Mohorovičić travel-time curve (a) (Courtesy of the Department of Geophysics, Faculty of Science, Zagreb) [26] and (b) best fitting curves

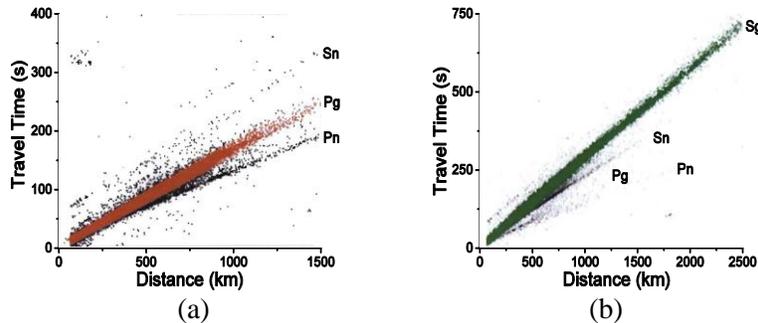

(a) (b)

FIG. 9. Travel time-distance plots of the Pg, Pn, Sg, Sn phases in Eurasia [27].

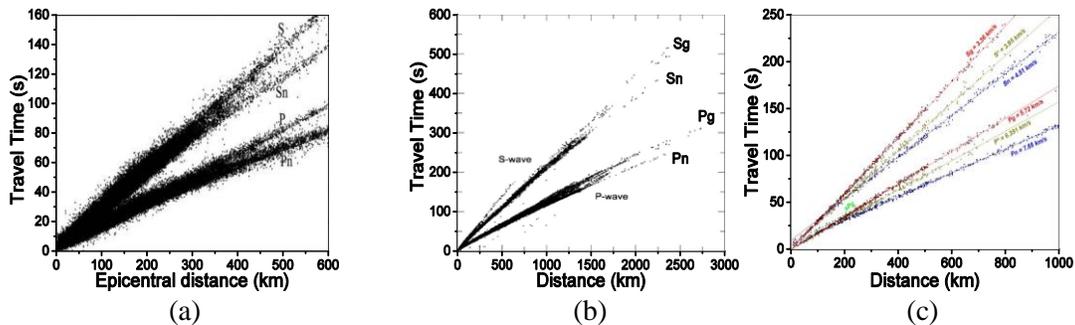

(a) (b) (c)

FIG. 10. Travel time-distance plot of the Pn, Pg, Sn and Sg phases in (a) Southwest China [28] and (b) Japan [29], and (c) eastern Russia [30]. In addition to the Pn, Pg, Sn, Sg phases, the $P^*$, and $S^*$ phases are also recorded from 92 Sakhalin earthquakes.

## Discussion

A new theoretical foundation for multiple scattering theories has been established to quantitatively modeling the coherent wave propagation in strong scattering materials. In a series of papers [13-14] the author systematically studied the scattering and attenuation characteristics of coherent waves in heterogeneous materials with equiaxed two-phase materials [13], equiaxed polycrystals [14], and tri-axial two-phase materials in this work. The renormalization technique for both isotropic and anisotropic materials is developed in detail. A number of new and important developments of the new model are foreseeable. First, we can develop the corresponding theory for polycrystalline materials with elongated grains and/or microscopic or macroscopic textures. These theories have significant applications in polycrystal microstructure and texture characterization [4-5]. Second, the renormalized technique can also be introduced to the development of the Bethe-Salpeter equation. It is noted that the coherent wave attenuates significantly or even disappear when the propagation distance and time lapse are large compared to the mean free path and mean free time, which is the case especially in strong scattering materials. The surviving incoherent, diffusive wavefield is most conveniently described in terms of energy transport. The radiative transport equation and diffusion equation can be further derived from the Bethe-Salpeter equation, which is anticipated to play a key role in the explanation of the coda wave attenuations. Third, the theory is of significant importance for ultrasonic microstructure characterization. Practical ultrasonic backscattering and transmitting measurements are normally performed by using a single or multiple focused or unfocused transducers in a water tank. Thus, the influence of the beam effects and the interfaces must be properly modeled and deconvolved to extract the response directly related to the microstructures. The new model enables us to develop new ultrasonic grain model and ultrasonic measurement model for strong scattering materials. It also paves the way for developing inversion method to reconstruct microstructures in polycrystal alloys.

One of the most striking predictions given by the new model is the bifurcation of the coherent waves at high frequencies and the new explanation to the formation of different seismic phases. It is already shown that the new model is able to give a consistent quantitative explanation to both the attenuation and wave fronts splitting recorded in real earthquakes based on a realistic lithosphere model. Here I would like to give more arguments to justify the rationality of the new explanations. Up to date, the Mohorovicic discontinuity is still out of human's reach although geophysicists allover the world have made great efforts to drilling deep holes into Earth. The deepest hole, named the Kola superdeep borehole, drilled by Soviet Union scientists is only about 12 km in depth, still far from the Moho (average depth of Moho is assumed to be 50 km). Therefore, there is no direct evidence to show the existence of the discontinuity. We should not misidentify rock boundaries and boundaries of any substructures exposed to Earth surface as the old Moho discontinuity. Actually, there are numerous "discontinuities" in the lithosphere. Moreover, we should not fully trust the seismic imaging methods such as traveltime tomography, which make use of only a specific arrival of the full seismic wave train. In addition, most imaging methods require an initial velocity model, which lead to model-dependent imaging results. The most trustworthy identification of the discontinuity relies on the development of exact 3D full-elastic inverse scattering theory. This dream is still far from being realized in the foreseeable future.

While the existence of the Moho discontinuity is challenged by the new theory, the author still keeps an open mind to all the possibilities. For any discontinuities to be claimed as the Moho or Conrad discontinuity, the following three conditions should be satisfied: 1) Waveform characteristics consistency, this means the waveform predicted by different imaging method should coincides with those observed in real earthquakes, includes the amplitude attenuation, coda, and the frequency composition; 2) Depth consistency, the depth of the discontinuity from imaging method should agree with the theoretical assumption; and 3) Mechanical Properties consistency, the density and elastic moduli, the velocity distribution above and below the discontinuity should satisfy the theoretical assumption. The current observation and imaging results does not support these discontinuities.

## Conclusion

The multiple scattering theory for the strong scattering heterogeneous elastic media with aligned tri-axial inhomogeneities are developed with the help of Feynman's diagram technique and renormalization method. The dispersion and attenuation of representative material models of the lithosphere are solved from the dispersion equation under the first-order-smoothing approximation. The new theory provides a consistent and quantitative explanation to the observed seismic attenuation and generation of different seismic phases. The model also provides a general theoretical framework for quantitative characterization and inversion of the statistical properties like characteristic size and aspect ratio of the non-equiaxed heterogeneities in Earth or polycrystalline alloys.

## Appendix A Calculation of the singularity tensor $S_{ij}$

In this appendix we discuss a method to calculate the singularity tensor of heterogeneous media with general tri-axial inhomogeneities. It has been pointed out that the explicit expression of Green's function in spatial domain is not available, so Method 1 introduced in [13] is invalid. In the frequency-wavenumber domain, the eigenvalues of the Christoffel tensor are dependent on the direction of the wavevector, as a result, integrals involving the dynamic Green's function in the frequency-wavenumber domain is also very difficult to evaluate. Method 2 in [13] is failed either. The only method to calculate the singularity is to use the static Green's tensor, i.e., Method 3 in [13]. In this method, the singularity of the Green tensor is extracted by eliminating the zero-frequency, i.e., static portion of the spectral domain convolution of the Green's function and the spectral correlation function. The poles of the spectral correlation function have an extremely simple form if we introduce a proper nonorthogonal ellipsoidal coordinate system

$$s_x = \frac{1}{a_x} s \sin\theta \cos\varphi, \quad s_y = \frac{1}{a_y} s \sin\theta \sin\phi, \quad s_z = \frac{1}{a_z} s \cos\theta, \tag{A1}$$

and the volume element is

$$d^3\mathbf{s} = ds_x \wedge ds_y \wedge ds_z = \frac{1}{a_x a_y a_z} s^2 \sin\theta \, ds \wedge d\theta \wedge d\varphi, \tag{A2}$$

In the new coordinate system, the denominator in the spectral correlation function becomes

$$1 + s_x^2 a^2 + s_y^2 b^2 + s_z^2 c^2 = 1 + s^2, \tag{A3}$$

Substitution of these results into Eq. (31) yields

$$S_{\alpha ij\beta} = \frac{1}{4\pi^2} \int_0^{+\infty} \frac{s^2}{(1+s^2)^2} ds \int_0^\pi \sin\theta \, d\theta \int_0^{2\pi} d\phi [s_j s_\beta \tilde{G}_{\alpha i}^0(\mathbf{s},\omega) + s_i s_\beta \tilde{G}_{\alpha j}^0(\mathbf{s},\omega) + s_j s_\alpha G_{\beta i}^0(\mathbf{s},\omega) + s_i s_\alpha G_{\beta j}^0(\mathbf{s},\omega)], \tag{A4}$$

The integral in Eq. (A4) has two poles of second order at $s = \pm i$. It can be evaluated using the residual theorem by taking the integral path shown in Fig. A1.

Introducing the compact index notation

$S_{11} = S_{1111}$, $S_{22} = S_{2222}$, $S_{33} = S_{3333}$, $S_{12} = S_{1221}$, $S_{13} = S_{1331}$, $S_{14} = S_{1231}$, $S_{15} = S_{1131}$, $S_{16} = S_{1121}$, $S_{23} = S_{2332}$, $S_{24} = S_{2232}$, $S_{25} = S_{2132}$,

$S_{26} = S_{2122}$, $S_{34} = S_{3233}$, $S_{35} = S_{3133}$, $S_{36} = S_{3123}$, $S_{44} = S_{2233}$, $S_{45} = S_{2133}$, $S_{46} = S_{2123}$, $S_{55} = S_{1133}$, $S_{56} = S_{1123}$, $S_{66} = S_{1122}$, (A5)

The singularity tensor can be written in matrix form

$$S_{ij} = \begin{bmatrix} S_{11} & S_{12} & S_{13} & S_{14} & S_{15} & S_{16} \\ S_{12} & S_{22} & S_{23} & S_{24} & S_{25} & S_{26} \\ S_{13} & S_{23} & S_{33} & S_{34} & S_{35} & S_{36} \\ S_{14} & S_{24} & S_{34} & S_{44} & S_{45} & S_{46} \\ S_{15} & S_{25} & S_{35} & S_{45} & S_{55} & S_{56} \\ S_{16} & S_{26} & S_{36} & S_{46} & S_{56} & S_{66} \end{bmatrix},$$ (A6)

For the tri-axial inhomogeneities, the singularity matrix has orthogonal symmetry as the reference medium,

$$S_{ij} = \begin{bmatrix} S_{11} & S_{12} & S_{13} & 0 & 0 & 0 \\ S_{12} & S_{22} & S_{23} & 0 & 0 & 0 \\ S_{13} & S_{23} & S_{33} & 0 & 0 & 0 \\ 0 & 0 & 0 & S_{44} & 0 & 0 \\ 0 & 0 & 0 & 0 & S_{55} & 0 \\ 0 & 0 & 0 & 0 & 0 & S_{66} \end{bmatrix},$$ (A7)

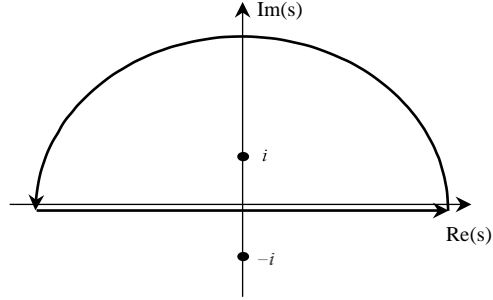

Figure A1. Integral path in the complex-s plane

The explicit expressions of the components are listed as follows

$$S_{11} = \frac{1}{4\pi} \int_0^\pi d\theta \int_0^{2\pi} d\phi \frac{1}{\Delta_s} \frac{1}{a_x^2} \sin^3\theta \cos^2\phi \left[ -\frac{(c_{23}+c_{44})^2}{a_y^2 a_z^2} \sin^2\theta \sin^2\phi \cos^2\theta + \right.$$
$$\left. \left( \frac{c_{66}}{a_x^2}\sin^2\theta\cos^2\phi + \frac{c_{22}}{a_y^2}\sin^2\theta\sin^2\phi + \frac{c_{44}}{a_z^2}\cos^2\theta \right) \left( \frac{c_{55}}{a_x^2}\sin^2\theta\cos^2\phi + \frac{c_{44}}{a_y^2}\sin^2\theta\sin^2\phi + \frac{c_{33}}{a_z^2}\cos^2\theta \right) \right],$$ (A8)

$$S_{22} = \frac{1}{4\pi} \int_0^\pi d\theta \int_0^{2\pi} d\phi \frac{1}{\Delta_s} \frac{1}{a_y^2} \sin^3\theta \sin^2\phi \left[ -\frac{(c_{13}+c_{55})^2}{a_x^2 a_z^2} \sin^2\theta \cos^2\phi \cos^2\theta + \right.$$
$$\left. \left( \frac{c_{11}}{a_x^2}\sin^2\theta\cos^2\phi + \frac{c_{66}}{a_y^2}\sin^2\theta\sin^2\phi + \frac{c_{55}}{a_z^2}\cos^2\theta \right) \left( \frac{c_{55}}{a_x^2}\sin^2\theta\cos^2\phi + \frac{c_{44}}{a_y^2}\sin^2\theta\sin^2\phi + \frac{c_{33}}{a_z^2}\cos^2\theta \right) \right],$$ (A9)

$$S_{33} = \frac{1}{4\pi} \int_0^\pi d\theta \int_0^{2\pi} d\phi \frac{1}{\Delta_s} \frac{1}{a_z^2} \sin\theta \cos^2\theta \left[ -\frac{(c_{12}+c_{66})^2}{a_x^2 a_y^2} \sin^4\theta \sin^2\phi \cos^2\phi + \right.$$
$$\left. \left( \frac{c_{11}}{a_x^2}\sin^2\theta\cos^2\phi + \frac{c_{66}}{a_y^2}\sin^2\theta\sin^2\phi + \frac{c_{55}}{a_z^2}\cos^2\theta \right) \left( \frac{c_{66}}{a_x^2}\sin^2\theta\cos^2\phi + \frac{c_{22}}{a_y^2}\sin^2\theta\sin^2\phi + \frac{c_{44}}{a_z^2}\cos^2\theta \right) \right],$$ (A10)

$$S_{12} = \frac{1}{4\pi}\int_0^\pi d\theta \int_0^{2\pi} d\phi \frac{1}{\Delta_s} \frac{1}{a_x a_y} \sin^3\theta \cos\phi \sin\phi \left[ \frac{(c_{13}+c_{55})(c_{23}+c_{44})}{a_x a_y a_z^2} \sin^2\theta \cos\phi \sin\phi \cos^2\theta - \right.$$
$$\left. \frac{c_{12}+c_{66}}{a_x a_y} \sin^2\theta \cos\phi \sin\phi \left( \frac{c_{55}}{a_x^2}\sin^2\theta\cos^2\phi + \frac{c_{44}}{a_y^2}\sin^2\theta\sin^2\phi + \frac{c_{33}}{a_z^2}\cos^2\theta \right) \right],$$
(A11)

$$S_{13} = \frac{1}{4\pi}\int_0^\pi d\theta \int_0^{2\pi} d\phi \frac{1}{\Delta_s} \frac{1}{a_x a_z} \sin^2\theta \cos\phi \cos\theta \left[ \frac{(c_{12}+c_{66})(c_{23}+c_{44})}{a_x a_z a_y^2} \sin^3\theta \cos\phi \sin^2\phi \cos\theta - \right.$$
$$\left. \frac{c_{13}+c_{55}}{a_x a_z} \sin\theta\cos\phi\cos\theta \left( \frac{c_{66}}{a_x^2}\sin^2\theta\cos^2\phi + \frac{c_{22}}{a_y^2}\sin^2\theta\sin^2\phi + \frac{c_{44}}{a_z^2}\cos^2\theta \right) \right],$$
(A12)

$$S_{23} = \frac{1}{4\pi}\int_0^\pi d\theta \int_0^{2\pi} d\phi \frac{1}{\Delta_s} \frac{1}{a_y a_z} \sin^2\theta \sin\phi \cos\theta \left[ \frac{(c_{12}+c_{66})(c_{13}+c_{55})}{a_y a_z a_x^2} \sin^3\theta \cos^2\phi \sin\phi \cos\theta - \right.$$
$$\left. \frac{c_{23}+c_{44}}{a_y a_z} \sin\theta\sin\phi\cos\theta \left( \frac{c_{11}}{a_x^2}\sin^2\theta\cos^2\phi + \frac{c_{66}}{a_y^2}\sin^2\theta\sin^2\phi + \frac{c_{55}}{a_z^2}\cos^2\theta \right) \right],$$
(A13)

$$S_{44} = \frac{1}{16\pi}\int_0^\pi d\theta \int_0^{2\pi} d\phi \frac{1}{\Delta_s} \left\{ \frac{1}{a_z^2}\sin\theta\cos^2\theta \left[ -\frac{(c_{13}+c_{55})^2}{a_x^2 a_z^2}\sin^2\theta\cos^2\phi\cos^2\theta + \left( \frac{c_{55}}{a_x^2}\sin^2\theta\cos^2\phi + \frac{c_{44}}{a_y^2}\sin^2\theta\sin^2\phi + \frac{c_{33}}{a_z^2}\cos^2\theta \right) \right. \right.$$
$$\left. \cdot \left( \frac{c_{11}}{a_x^2}\sin^2\theta\cos^2\phi + \frac{c_{66}}{a_y^2}\sin^2\theta\sin^2\phi + \frac{c_{55}}{a_z^2}\cos^2\theta \right) \right] + \frac{1}{a_y^2}\sin^3\theta\sin^2\phi \left[ -\frac{(c_{12}+c_{66})^2}{a_x^2 a_y^2}\sin^4\theta\cos^2\phi\sin^2\phi \right.$$
$$\left. + \left( \frac{c_{66}}{a_x^2}\sin^2\theta\cos^2\phi + \frac{c_{22}}{a_y^2}\sin^2\theta\sin^2\phi + \frac{c_{44}}{a_z^2}\cos^2\theta \right)\left( \frac{c_{11}}{a_x^2}\sin^2\theta\cos^2\phi + \frac{c_{66}}{a_y^2}\sin^2\theta\sin^2\phi + \frac{c_{55}}{a_z^2}\cos^2\theta \right) \right]$$
$$+ \frac{2}{a_y a_z}\sin^2\theta\sin\phi\cos\theta \left[ \frac{(c_{12}+c_{66})(c_{13}+c_{55})}{a_y a_z a_x^2}\sin^3\theta\cos^2\phi\sin\phi\cos\theta \right.$$
$$\left. \left. - \frac{c_{23}+c_{44}}{a_y a_z}\sin\theta\sin\phi\cos\theta \left( \frac{c_{11}}{a_x^2}\sin^2\theta\cos^2\phi + \frac{c_{66}}{a_y^2}\sin^2\theta\sin^2\phi + \frac{c_{55}}{a_z^2}\cos^2\theta \right) \right] \right\},$$
(A14)

$$S_{55} = \frac{1}{16\pi}\int_0^\pi d\theta \int_0^{2\pi} d\phi \frac{1}{\Delta_s} \left\{ \frac{1}{a_z^2}\sin\theta\cos^2\theta \left[ -\frac{(c_{23}+c_{44})^2}{a_y^2 a_z^2}\sin^2\theta\sin^2\phi\cos^2\theta + \left( \frac{c_{55}}{a_x^2}\sin^2\theta\cos^2\phi + \frac{c_{44}}{a_y^2}\sin^2\theta\sin^2\phi + \frac{c_{33}}{a_z^2}\cos^2\theta \right) \right. \right.$$
$$\left. \cdot \left( \frac{c_{66}}{a_x^2}\sin^2\theta\cos^2\phi + \frac{c_{22}}{a_y^2}\sin^2\theta\sin^2\phi + \frac{c_{44}}{a_z^2}\cos^2\theta \right) \right] + \frac{1}{a_x^2}\sin^3\theta\cos^2\phi \left[ -\frac{(c_{12}+c_{66})^2}{a_x^2 a_y^2}\sin^4\theta\cos^2\phi\sin^2\phi \right.$$
$$\left. + \left( \frac{c_{66}}{a_x^2}\sin^2\theta\cos^2\phi + \frac{c_{22}}{a_y^2}\sin^2\theta\sin^2\phi + \frac{c_{44}}{a_z^2}\cos^2\theta \right)\left( \frac{c_{11}}{a_x^2}\sin^2\theta\cos^2\phi + \frac{c_{66}}{a_y^2}\sin^2\theta\sin^2\phi + \frac{c_{55}}{a_z^2}\cos^2\theta \right) \right]$$
$$+ \frac{2}{a_x a_z}\sin^2\theta\cos\phi\cos\theta \left[ \frac{(c_{12}+c_{66})(c_{23}+c_{44})}{a_x a_z a_y^2}\sin^3\theta\cos\phi\sin^2\phi\cos\theta \right.$$
$$\left. \left. - \frac{c_{13}+c_{55}}{a_x a_z}\sin\theta\cos\phi\cos\theta \left( \frac{c_{66}}{a_x^2}\sin^2\theta\cos^2\phi + \frac{c_{22}}{a_y^2}\sin^2\theta\sin^2\phi + \frac{c_{44}}{a_z^2}\cos^2\theta \right) \right] \right\},$$
(A15)

$$S_{66} = \frac{1}{16\pi}\int_0^\pi d\theta \int_0^{2\pi} d\phi \frac{1}{\Delta_s} \left\{ \frac{1}{a_y^2}\sin^3\theta\sin^2\phi \left[ -\frac{(c_{23}+c_{44})^2}{a_y^2 a_z^2}\sin^2\theta\sin^2\phi\cos^2\theta + \left( \frac{c_{55}}{a_x^2}\sin^2\theta\cos^2\phi + \frac{c_{44}}{a_y^2}\sin^2\theta\sin^2\phi + \frac{c_{33}}{a_z^2}\cos^2\theta \right) \right. \right.$$
$$\left. \cdot \left( \frac{c_{66}}{a_x^2}\sin^2\theta\cos^2\phi + \frac{c_{22}}{a_y^2}\sin^2\theta\sin^2\phi + \frac{c_{44}}{a_z^2}\cos^2\theta \right) \right] + \frac{1}{a_x^2}\sin^3\theta\cos^2\phi \left[ -\frac{(c_{13}+c_{55})^2}{a_x^2 a_z^2}\sin^2\theta\cos^2\phi\cos^2\theta \right.$$
$$\left. + \left( \frac{c_{55}}{a_x^2}\sin^2\theta\cos^2\phi + \frac{c_{44}}{a_y^2}\sin^2\theta\sin^2\phi + \frac{c_{33}}{a_z^2}\cos^2\theta \right)\left( \frac{c_{11}}{a_x^2}\sin^2\theta\cos^2\phi + \frac{c_{66}}{a_y^2}\sin^2\theta\sin^2\phi + \frac{c_{55}}{a_z^2}\cos^2\theta \right) \right]$$
$$+ \frac{2}{a_x a_y}\sin^3\theta\cos\phi\sin\phi \left[ \frac{(c_{13}+c_{55})(c_{23}+c_{44})}{a_x a_y a_z^2}\sin^2\theta\cos\phi\sin\phi\cos^2\theta \right.$$
$$\left. \left. - \frac{c_{12}+c_{66}}{a_x a_y}\sin^2\theta\cos\phi\sin\phi \left( \frac{c_{55}}{a_x^2}\sin^2\theta\cos^2\phi + \frac{c_{44}}{a_y^2}\sin^2\theta\sin^2\phi + \frac{c_{33}}{a_z^2}\cos^2\theta \right) \right] \right\},$$
(A16)

where

$$\begin{aligned}\Delta_s =& \left(\frac{c_{11}}{a_x^2}\sin^2\theta\cos^2\phi + \frac{c_{66}}{a_y^2}\sin^2\theta\sin^2\phi + \frac{c_{55}}{a_z^2}\cos^2\theta\right)\left(\frac{c_{66}}{a_x^2}\sin^2\theta\cos^2\phi + \frac{c_{22}}{a_y^2}\sin^2\theta\sin^2\phi + \frac{c_{44}}{a_z^2}\cos^2\theta\right)\cdot \\ & \left(\frac{c_{55}}{a_x^2}\sin^2\theta\cos^2\phi + \frac{c_{44}}{a_y^2}\sin^2\theta\sin^2\phi + \frac{c_{33}}{a_z^2}\cos^2\theta\right) \\ & - \left(\frac{c_{11}}{a_x^2}\sin^2\theta\cos^2\phi + \frac{c_{66}}{a_y^2}\sin^2\theta\sin^2\phi + \frac{c_{55}}{a_z^2}\cos^2\theta\right)\frac{(c_{23}+c_{44})^2}{a_y^2 a_z^2}\sin^2\theta\sin^2\phi\cos^2\theta \\ & - \left(\frac{c_{55}}{a_x^2}\sin^2\theta\cos^2\phi + \frac{c_{44}}{a_y^2}\sin^2\theta\sin^2\phi + \frac{c_{33}}{a_z^2}\cos^2\theta\right)\frac{(c_{12}+c_{66})^2}{a_x^2 a_y^2}\sin^4\theta\sin^2\phi\cos^2\phi \\ & - \left(\frac{c_{66}}{a_x^2}\sin^2\theta\cos^2\phi + \frac{c_{22}}{a_y^2}\sin^2\theta\sin^2\phi + \frac{c_{44}}{a_z^2}\cos^2\theta\right)\frac{(c_{13}+c_{55})^2}{a_x^2 a_z^2}\sin^2\theta\cos^2\phi\cos^2\theta \\ & + 2\frac{(c_{12}+c_{66})(c_{13}+c_{55})(c_{23}+c_{44})}{a_x^2 a_y^2 a_z^2}\sin^4\theta\sin^2\phi\cos^2\phi\cos^2\theta.\end{aligned} \quad (A17)$$